\documentclass[final,3p,times,authoryear]{elsarticle}

\usepackage{amssymb, amsmath, amsthm, lineno, epsfig, bm, latexsym}
\usepackage{pdflscape}
\usepackage{natbib}

\newtheorem{remark}{Remark}

\date{}
\usepackage[dvipsnames]{xcolor}
\definecolor{mygray}{gray}{0.98}
\usepackage[framemethod=TikZ]{mdframed}
\usepackage{lipsum}
\mdfdefinestyle{MyFrame}{%
	linecolor=black,
	outerlinewidth=0.003cm,
	roundcorner=18pt,
	innertopmargin=\baselineskip,
	innerbottommargin=\baselineskip,
	backgroundcolor=mygray}

\begin{document}
	\begin{frontmatter}
		\title{Mixture of linear experts model for censored data:\\
			A novel approach with scale-mixture of normal distributions}
		\author{Elham Mirfarah\fnref{label1}\corref{cor1}}
		
		\author{Mehrdad Naderi\fnref{label1}}
		%
		\author{Ding-Geng Chen \fnref{label1,label2}}
		
		\cortext[cor1]{Corresponding author\fnref{cor1}}
		\address[label1]{Department of Statistics, Faculty of Natural \& Agricultural Sciences,
			University of Pretoria, Pretoria, South Africa}
		\address[label2]{Department of Biostatistics, University of North Carolina, Chapel Hill, NC 27599, USA}
		\begin{abstract}
			The classical mixture of linear experts (MoE) model is one of the widespread statistical frameworks for modeling, 
			classification, and clustering of data. Built on the normality assumption of the error terms for mathematical and
			computational convenience, the classical MoE model has two challenges: 1) it is sensitive to atypical observations
			and outliers, and 2) it might produce misleading inferential results for censored data. The paper is then aimed
			to resolve these two challenges, simultaneously, by proposing a novel robust MoE model for model-based clustering and discriminant 
			censored data with the scale-mixture of normal class of distributions for the unobserved error terms. Based 
			on this novel model, we develop an analytical expectation-maximization (EM) type algorithm to obtain the maximum 
			likelihood parameter estimates. Simulation studies are carried out to examine the performance, effectiveness, 
			and robustness of the proposed methodology. Finally, real data is used to illustrate the superiority of the new model.
		\end{abstract}
		\begin{keyword}
			Mixture of linear experts model\sep Scale-mixture of normal class of distributions \sep EM-type algorithm\sep Censored data
		\end{keyword}
	\end{frontmatter}
	
	\section{Introduction} \label{Intro}	
	The issue of model-based clustering has recently received considerable attention in statistics with applications in medical
	sciences, public health and engineering as shown in \cite{feigelson2012,wang2019,shafiei2020}. The grouping structure identification in the data 
	usually provides informative results for solving real-world problems. One of the most acknowledged statistical tools for model-based
	clustering is the finite mixture (FM) model. The FM model, initially introduced by \cite{redner1984},
	is a convex linear combination of the probability density functions (pdf) given by
	\[	f(y;\bm\Theta) = \sum_{j=1}^G \pi_j\; g(y;\bm\theta_j),\]
	where $G$ is the total number of clusters,  $\pi_j$'s are the mixing proportions subjected to $\sum_{j=1}^G \pi_j =1$, 
	$g(\cdot;\bm\theta_j)$ is the associated pdf of the $j$th underlying mixing component parametrized by $\bm\theta_j$, and $\bm\Theta=(\bm\theta_1,\ldots,\bm\theta_G,\pi_1,\ldots,\pi_{G-1})$. 
	The mixing proportion is, in fact, $\pi_j=Pr(Z^\ast=j)$ where the hidden categorical random variable $Z^*$ indicates from which
	component each observation is arisen. Upon the normality and/or non-normality assumptions 
	for the mixing components, various FM models have recently been introduced for modeling heterogeneous data. They have widely 
	been employed in scientific studies such as genetics, image processing, medicine, economics and astronomy, for example in \cite{Wang2018, 
		Punzo2018, Sugasawa2018, Naderi2017a, Naderi2019, Tomarchio2019, Morris2019} to name a few recently published papers.
	
	In the context of regression analysis, the FM models are also found appealing applications to investigate the relationship 
	between the random phenomena under study arisen from various unknown latent homogeneous groups. Specifically, 
	the FM regression model relies on the assumption that the pdf of underlying distribution is
	\[
	f(y;\bm\Theta)=\sum_{j=1}^G\pi_j g(y-\bm x^\top\bm\beta_j;\bm\theta_j),
	\]
	where $\bm x=(1,x_1,\ldots,x_{p-1})^\top\in \mathbb{R}^p$ is the vector of $p$ explanatory variables corresponding to response $y$, and $\bm\beta_j=(\beta_{j0},\ldots,\beta_{j(p-1)})^\top$, denotes the regression coefficients of the $j$th component.	
	In this regard, \cite{liu2014} proposed the skew-normal mixture
	regression model by considering the pdf of skew-normal distribution as the mixing component and applied it 
	to the physiological data 
	to illustrate its utility. 
	\cite{Hu2017} introduced FM regression model by assuming that the components have log-concave error densities
	and developed two EM-type \citep{dempster1977} algorithms to obtain the maximum likelihood (ML) parameter estimates.
	\cite{Lamont2016} also investigated the effect of modeling covariance between independent variables and
	latent classes on the fitting/clustering results.
	
	Built up from this FM regression model, the MoE model \citep{jacobs1991} is perhaps one of the most acknowledged
	approaches in the statistics and machine learning fields. Although the MoE and FM regression models share similar structure, they differ in many aspects. In formulation of the MoE model, it is assumed that both mixing
	proportions and component densities conditionally depend on some input covariates. More precisely, let
	$Y\in \mathbb{R}$ be the response variable, $\bm x\in \mathbb{R}^p$ and $\bm r=(1,r_1,\ldots,r_{q-1})^\top\in \mathbb{R}^q$  are
	the vector of explanatory and covariate values corresponding to $Y$. Instead of considering constant mixing
	component in FM regression model, the MoE model assumes that $\pi_j$ to be modeled as a function (generally logistic or
	softmax function) of input $\bm r$, known as a gating function. For instance, the pdf of the normal-based MoE (MoE-N) is
	\begin{equation}\label{NMoE}
		f(y;\bm\Theta)= \sum_{j=1}^{G} \pi_j(\bm r;\bm\tau) \phi(y;\bm x^\top\bm\beta_j, \sigma_j^2),
	\end{equation}
	where $\phi(\cdot,\mu,\sigma^2)$ is the pdf of normal distribution with the location and scale parameters
	$\mu$ and $\sigma^2$, $\mathcal{N}(\mu,\sigma^2)$, for gating parameters $\bm\tau=(\bm\tau_1^\top,\ldots,\bm\tau_{G-1}^\top)^\top$
	with $\bm\tau_j=(\tau_{j0},\ldots,\tau_{j(q-1)})^\top$,
	\begin{equation}\label{MixPro}
	\pi_{j}(\bm r;\bm\tau)= Pr(Z^*=j|\bm r) =\frac{\exp\{\bm\tau_j^\top \bm r \}}{1+\sum_{l=1}^{G-1}\exp\{\bm\tau_l^\top \bm r\}},
	\end{equation}
	and for $\bm\theta_j=(\bm\beta_j,\sigma^2_j)$, the model parameters set is $\bm\Theta=(\bm\theta_1,\ldots,\bm\theta_G,\bm\tau)$.
	It should be emphasized that $\bm x$ and $\bm r$ can be exactly or partially identical.
	Since the introduction of the MoE-N model, considerable amount of contributions have been produced
	to overcome its potential deficiency in analyzing skew and heavy-tail distributed data. See for instance the
	works by \cite{Nguyen2016,Chamroukhi2016, Chamroukhi2017} on proposing the Laplace, Student-$t$ and
	skew-$t$ MoE models, respectively. 
	
	In many practical situations, such as economic and clinical studies, medical research and epidemiological
	cancer studies, the data are collected under some imposed detection limits. It might lead to incomplete 
	data with different types of interval, left and/or right-censored responses. In this regard, 
	censored regression model with the normality assumption for the error terms, known as Tobit model, was 
	constructed by \cite{tobin1958}. Since then, the extensions of Tobit model have been introduced by researchers 
	to draw robust inference from censored data. For instance, using the scale-mixture of normal (SMN) class of distributions 
	for the error terms, \cite{garay2016, garay2017} presented the nonlinear and linear censored regression models
	to overcome the problem of atypical observations in the data. \cite{mattos2018} also proposed censored
	linear regression model with the scale-mixture of skew-normal class of distributions to accommodate asymmetrically
	distributed censored datasets. Moreover, mixture of censored regression models based on the Student-$t$ model and on the 
	SMN class of distributions were proposed by \cite{lachos2019, Zeller2018} as a flexible
	approach for modeling multimodal censored data with fat tails.
	
	Extending the proven proficiency of the MoE model in statistical applications, the main objective of the
	current paper is to propose a MoE model based on the SMN class of distributions for censored data, hereafter 
	referred as ``MoE-SMN-CR model". Due to the computational complexity, we develop an innovative EM-type algorithm to obtain the ML parameter estimates. The associated variance-covariance matrix of the ML estimators 
	is also approximated by an information-based approach. To illustrate the computational aspects
	and practical performance of the proposed methodology, a real data analysis and several simulation studies are
	presented.
	
	The remainder of the paper is organized as follows. Section \ref{sec.2} briefly reviews the SMN class
	of distributions. Model formulation and parameter estimation procedure of the MoE-SMN-CR model are
	presented in Section \ref{sec.3}. Four simulation studies are conducted in Section \ref{sec4} to check the 
	asymptotic properties of the ML estimates as well as to investigate the performance of the proposed model. 
	The applicability of the proposed method is illustrated in Section \ref{sec5} by analyzing wage-rates dataset. 
	Finally, we conclude the paper with a discussion and suggestions for future work in Section \ref{sec6}.
	
	\section{An overview on the scale-mixture of normal class of distributions}\label{sec.2}
	A random variable $Y$ follows an scale-mixture of normal (SMN) distribution, denoted by $\mathcal{SMN}(\mu, \sigma^2,\bm\nu)$,
	if it is generated by
	\begin{equation}\label{SMN-lin-rep}
		Y=\mu+U^{-1/2}V,\qquad V\perp U,
	\end{equation}
	where $V\sim \mathcal{N}(0,\sigma^2)$, $U$ (scale mixture factor) is a positive random variable with the cumulative
	distribution function (cdf) $H(\cdot;\bm\nu)$, and the symbol $\perp$ indicates independence.
	Referring to \eqref{SMN-lin-rep}, the hierarchical representation of the SMN class of distributions can be written as
	\begin{equation}\label{SMN-hir-rep}
		Y|U=u\sim N(\mu, u^{-1}\sigma^2), \qquad U\sim H(u; \bm\nu),
	\end{equation}
	accordingly, the pdf of the random variable $Y$ is obtained as
	\[	f_{_\text{SMN}}(y;\mu, \sigma^2,\bm\nu) = \int_{0}^\infty \phi(y;\mu, u^{-1}\sigma^2)\; dH(u; \bm\nu),
		\qquad y\in\mathbb{R}.\]
	In what follows, $f_{_\text{SMN}}(\cdot;\bm\nu)$ and $F_{_\text{SMN}}(\cdot;\bm\nu)$ will be used to denote the
	pdf and cdf of the standard SMN distribution ($\mu=0,\sigma^2=1$). With different specifications of the distribution of $U$, 
	many special cases of the general SMN class of distributions can be obtained. We focus on a few commonly used
	examples of the SMN class of distributions in this paper:
	
	\begin{itemize}
		\item Normal (N) distribution: The SMN class of distributions contains the normal model as $U=1$ with
		probability one.
		
		\item Student-$t$ (T) distribution: If $U\sim Gamma ({\nu}/{2},{\nu}/{2})$, where $Gamma(\alpha, \beta)$ represents
		the  gamma distribution with shape and scale parameters $\alpha$ and $\beta$, respectively, the
		random variable $Y$  then follows the Student-$t$ distribution, $Y\sim \mathcal{T}(\mu,\sigma^2,\nu)$. For $\nu =1$
		the Student-$t$ distribution turns into the Cauchy distribution which has no defined mean and variance.
		
		\item Slash (SL) distribution: Let $U$ in \eqref{SMN-lin-rep} follows $Beta(\nu,1)$, where $Beta(a,b)$ signifies
		the beta distribution with parameter $a$ and $b$. Then, $Y$ distributed as a slash model, denoted by $Y\sim
		\mathcal{SL}(\mu,\sigma^2,\nu)$, with pdf
		\[
		f_{_\text{SL}}(y;\mu,\sigma^2,\nu) = \nu \int_{0}^1 u^{\nu-1}\phi(y;\mu, u^{-1}\sigma^2)\; du,
		\qquad y\in\mathbb{R}.
		\]
		
		\item Contaminated-normal (CN) distribution: Let $U$ be a discrete random variable with pdf
		\[
		h(u;\nu,\gamma)= \nu \mathbb{I}_\gamma(u)+(1-\nu)\mathbb{I}_1(u), \qquad \nu,\gamma\in (0,1),
		\]
		where $\mathbb{I}_A(\cdot)$ represents the indicator function of the set $A$. The random variable $Y$ in
		\eqref{SMN-lin-rep} then follows the contaminated-normal distribution, $Y\sim \mathcal{CN}(\mu,\sigma^2,\nu,\gamma)$,
		which has the pdf
		\[
		f_{_\text{CN}}(y;\mu,\sigma^2,\nu,\gamma)= \nu \phi(y; \mu,\gamma^{-1}\sigma^2)+(1-\nu)\phi(y; \mu,\sigma^2),
		\qquad y\in\mathbb{R}.
		\]
		Note that in the pdf of CN distribution,  the parameter $\nu$ denotes the proportion of outliers (bad points)
		and $\gamma$ is the contamination factor.
	\end{itemize}
	
	More technical details and information of the SMN distribution family, used for the calculation of
	some conditional expectations involved in the proposed EM-type algorithm, are provided in the \ref{appa} with proof in
	\cite{garay2017}.
	We will refer to the MoE model of censored data based on the special cases of the SMN class of distributions as 
	MoE-N-CR, MoE-T-CR, MoE-SL-CR and MoE-CN-CR for the normal, Student-$t$, slash and contaminated-normal cases, respectively.
	
	\section{The scale-mixture of normal censored mixture of linear experts model}\label{sec.3}
	
	\subsection{Model specification}\label{sub.3.1}
	Extending the classical MoE model with normal distribution in model \eqref{NMoE}, 
	we consider the expert components formulated by the SMN class of distributions. 
	Therefore, the resulting pdf, in which the polynomial regression and multinomial logistic model are used
	for the components and mixing proportions, can be defined as
	\begin{equation}\label{MoE-SMN}
		f(y_i;\bm\Theta)= \sum_{j=1}^{G} \pi_j(\bm r_i;\bm\tau) f_{_\text{SMN}}(y_i;\bm x_i^\top\bm\beta_j, \sigma_j^2,\bm\nu_j),
		\quad \quad i=1,\ldots,n,
	\end{equation}
	where $\bm Y=(Y_1,\ldots,Y_n)^\top$ is the vector of response variables,
	$\bm{x}_i$ and $\bm{r}_i$ are the vector of explanatory and covariate variables corresponding to $Y_i$, 
	$\pi_{j}(\cdot;\bm\tau)$ is defined in \eqref{MixPro}, 
	and for $\bm\theta_j=(\bm\beta_j,\sigma^2_j,\bm\nu_j)$ the model parameters is $\bm\Theta=(\bm\theta_1,\ldots,\bm\theta_G,\bm\tau)$.
	
	In the MoE-SMN-CR model, we assume that the response variables are partially observed. In other word, we suppose some of the
	response variables are suffering from a type of censoring, that could be interval-, left- or right-censoring. 
	Thus, let the available response variable $Y_i$ be presented as the joint variables  
	$(W_i,\rho_i)$ where $W_i$ represents the uncensored reading $(W_i = Y_{Oi})$ or interval-censoring 
	$(W_i = (C_{i1},C_{i2}))$ and $\rho_i$ is the censoring indicator:
	$\rho_i=1$ if $C_{i1} \leq Y_i \leq C_{i2} $  and $\rho_i=0$ if $Y_i =Y_{Oi}$.
	Note that in this setting if $C_{i1} =- \infty$ (or $C_{i2}=+\infty$) the left-censoring (or right-censoring) is occurred and in the case $-\infty \neq C_{i1} < C_{i2} \neq +\infty$ the interval-censored realization is observed. We establish our methodology based on the interval-censoring scheme,
	however, the left/right-censoring schemes are also investigated in the simulation and real-data analyses.
	
	The aforementioned setting leads to divide $\bm Y$ to the sets of observed responses and
	censored cases. Hence, $\bm Y$ can be viewed as the latent variable since it is partially unobserved.
	Under these assumptions, the log-likelihood function of the MoE-SMN-CR model can be written as
	\begin{eqnarray} \label{loglike}
		\ell(\bm\Theta|\bm w, \bm\rho) = \sum_{i=1}^n \log \sum_{j=1}^G \pi_j(\bm r_i;\bm\tau)\left[{\sigma_j}^{-1}
		f_{_\text{SMN}}\left(\frac{w_i-\bm x_i^\top\bm\beta_j}{\sigma_j};\bm\nu_j\right)\right]^{1-\rho_i} 
		\left[ F_{_\text{SMN}}\left(\frac{c_{i2}-\bm x_i^\top\bm\beta_j}{\sigma_j}; \bm\nu_j\right) - 
		 F_{_\text{SMN}}\left(\frac{c_{i1}-\bm x_i^\top\bm\beta_j}{\sigma_j}; \bm\nu_j\right)\right]^{\rho_i} ,
	\end{eqnarray}
	where $\bm w=(w_1,\ldots,w_n)^\top$ and $\bm\rho = (\rho_1,\ldots,\rho_n)^\top$ denote the realizations of
	$\bm W=(W_1,\ldots,W_n)^\top$ and $\bm\rho = (\rho_1,\ldots,\rho_n)^\top$, respectively.
	
Due to complexity of the log-likelihood \eqref{loglike}, there is no analytical solution to obtain the ML estimate of parameters and therefore a numerical search algorithm should be developed. 
With the embedded hierarchical representation \eqref{SMN-hir-rep}, an innovative EM-type algorithm is developed to obtain the ML estimate for the MoE-SMN-CR model. 	

	\subsection{EM-based maximum likelihood parameter estimation}\label{ECMsec}
	Starting from \eqref{MoE-SMN} and defining the component label vector $\bm{Z}_{i}=(Z_{i1},\ldots,Z_{iG})^\top$
	in such a way that the binary latent component-indicators $Z_{ij}=1$ if and only if $Z_i^*=j$, we have
	\[
	Y_i |{Z}_{ij}=1 \sim \mathcal{SMN}(\bm x_i^\top\bm\beta_j, \sigma_j^2,\bm\nu_j), \quad \quad i=1,\ldots,n.
	\]
	Now using \eqref{SMN-hir-rep}, the hierarchical representation of the MoE-SMN-CR model is
	\begin{align*}
		Y_i |(\bm{x}_i, U=u_i, {Z}_{ij}=1) &\sim \mathcal{N}(\bm x_i^\top\bm\beta_j, u_i^{-1}\sigma_j^2), \\
		U_i| Z_{ij}=1 &\sim H(\cdot;\bm\nu_j),\\
		Z_i | \bm r_i &\sim \mathcal{M} \left( 1; \pi_1(\bm r_i,\bm\tau),\ldots, \pi_G(\bm r_i,\bm \tau)\right).
	\end{align*}
	where $\mathcal{M}(1;\cdot)$ denotes the one trail multinomial distribution.
	For the realizations $\bm y=(y_1,\ldots,y_n)^\top$, $\bm Z=(\bm Z_1^\top,\ldots,\bm Z_n^\top)^\top$ and the latent values
	$\bm u=(u_1,\ldots,u_n)^\top$, the log-likelihood function for $\bm \Theta$ associated with complete data
	$\bm y_c=(\bm w^\top,\bm\rho^\top,\bm y^\top, \bm u^\top,\bm Z^\top)^\top$, is therefore given by
	\begin{equation}\label{log-com}
		\ell_c(\bm\Theta|\bm y_c)= c + \sum_{i=1}^n \sum_{j=1}^G Z_{ij} \left\lbrace \log \pi_j(\bm r_i;\bm\tau)- \frac{1}{2}\log \sigma_j^2
		- \frac{u_{i}}{2\sigma_j^2} (y_i-\bm x_i^\top\bm\beta_j)^2 +\log h(u_i;\bm \nu_j)\right\rbrace,
	\end{equation}
	where $h(\cdot;\bm \nu_j)$ is the pdf of $U_i| Z_{ij}=1$ and $c$ is an additive constant.

We then develop an expectation conditional maximization either
(ECME; \cite{liu1994}) algorithm to estimate parameters from the MoE-SMN-CR model. 
The ECME algorithm is an extension of expectation conditional maximization
	(ECM; \cite{meng1993}) that not only inherits its stable properties (e.g. monotone convergence and implementation simplicity)
	but also can be faster than ECM. The iterative ECME algorithm replaces some CM-steps of the ECM with the CML-steps that
	maximize the corresponding contained log-likelihood function instead. The ECME algorithm for ML estimation of the MoE-SMN-CR
	model proceeds as follows:
	
	\begin{itemize}
		\item {\bf Initialization:}
		Set the number of iteration to $k = 0$ and choose a relative starting point. 
		Due to the multimodal  log-likelihood function in the FM and MoE models, the EM-type algorithm for
		obtaining parameter estimates might not give the global estimates if the initial points depart too far from the real values.
		Therefore, the choice of initialization process for the EM-based algorithms constitutes an fundamental issue.
		\cite{Nguyen2016} suggested the starting points for the Laplace MoE model via a modified version of the randomized
		initial assignment method \citep{mclachlan2004}. However, we recommend the following straightforward steps 
		for obtaining the starting points of the MoE-SMN-CR model.
		\begin{itemize}
			\item[(i)] Partition the sample into $G$ groups using either $K$-means clustering algorithm \citep{hartigan1979},
			$k$-medoids \citep{kaufman1990} or trim-$k$-means \citep{cuesta1997} methods.
			
			\item[(ii)] To initialize $\bm\tau^{(0)}$, two strategies can be adopted. As the first and simplest strategy,
			one can set $\bm\tau^{(0)} = \bm 0$. We note that by using this setting, the MoE model reduces to the
			FM regression model as a special case. In the second strategy, the information of grouping indices obtained from (i) can be used
			for initializing $\bm\tau$. Based on the grouping indices, one can fit the generalized linear model to the data
			and compute $\bm\tau^{(0)}$. 
			
			\item[(iii)] By utilizing the grouping indices of (i), the least squares method is applied to the $j$th group
			to obtain $\bm\beta_j^{(0)}$. Moreover, the standard deviation of residuals is used to initialize $\sigma^{(0)}$.
			\item[(iv)] Since the normal model belongs to the SMN class of distributions, we adapt
			$\bm\nu_j^{(0)}$ corresponds to an initial assumption near normality. For instance, we set $\nu_j=20$ in the
			MoE-T-CR and MoE-SL-CR models.		
		\end{itemize}

		\item {\bf E-Step:} At the iteration $k$, the expected value of the complete-data log-likelihood function
		\eqref{log-com}, known as the $Q$-function, is 
		\begin{align}\label{qfunc}
			Q(\bm\Theta| \bm\Theta^{(k)})
			=&  \sum_{i=1}^n \sum_{j=1}^G \hat{z}_{ij}^{(k)}\left\lbrace
			\log \pi_j(\bm r_i;\bm\tau) -\frac{1}{2}\log \sigma_j^2 -\frac{1}{2\sigma_j^2} \left( \widehat{uy^2}_{ij}^{(k)}+
			(\bm x_i^\top\bm\beta_j)^2 \hat{u}_{ij}^{(k)} -2  \widehat{uy}_{ij}^{(k)} \bm x_i^\top\bm\beta_j\right)
			+\hat{\Upsilon}_{ij}^{(k)}\right\rbrace,
		\end{align}
		where $\hat{z}_{ij}^{(k)} = E(Z_{ij}| w_i, \rho_i, \hat{\bm\theta}_j^{(k)})$,
		$\widehat{uy^2}_{ij}^{(k)} = E(U_{i}Y_i^2|  w_i, \rho_i, \hat{\bm\theta}_j^{(k)} ),$
		$\hat{u}_{ij}^{(k)}  = E(U_{i} |  w_i, \rho_i, \hat{\bm\theta}_j^{(k)} ),$ 
		$\widehat{uy}_{ij}^{(k)} = E(U_{i}Y_i |  w_i, \rho_i, \hat{\bm\theta}_j^{(k)}),$ and
		$\hat{\Upsilon}_{ij}^{(k)} = E\big(\log h(U_i;\bm\nu_j)|w_i,\rho_i, \hat{\bm\theta}_j^{(k)}\big)$.
		In what follows, we discuss about the computation of conditional expectations for both uncensored and censored
		cases. 		
		\begin{itemize}
		\item[(i)] For the uncensored observations, we have $\rho_i =0$ and so,
		$\hat{u}_{ij}^{(k)}  = E(U_{i} |  Y=y_i, \hat{\bm\theta}_j^{(k)})$,
		$\widehat{uy}_{ij}^{(k)} = y_i\hat{u}_{ij}^{(k)},$
		$\widehat{uy^2}_{ij}^{(k)} = y_i^2 \hat{u}_{ij}^{(k)},$		
		\[
			\hat{z}_{ij}^{(k)} =\frac{\pi_j(\bm r_i; \hat{\bm\tau}^{(k)})
				f_{_\text{SMN}}\left( y_i;\bm x_i^\top \hat{\bm\beta}_j^{(k)},\hat{\sigma}_j^{2(k)},\hat{\bm\nu}_j^{(k)}\right) }{
				\sum_{l=1}^G \pi_l(\bm r_i; \hat{\bm\tau}^{(k)})f_{_\text{SMN}}\left(y_i;\bm x_i^\top \hat{\bm\beta}_l^{(k)},\hat{\sigma}_l^{2(k)},\hat{\bm\nu}_l^{(k)}\right)},\qquad
			\hat{\Upsilon}_{ij}^{(k)} = E\big(\log h(U_i;\bm\nu_j)|Y=y_i, \hat{\bm\theta}_j^{(k)} \big).
		\]
		
		\item[(ii)] For the censored case which is $\rho_i =1$, we have
		\begin{align*}
			\hat{z}_{ij}^{(k)} &= E(Z_{ij}| c_{i1} \leq Y_i \leq c_{i2},\hat{\bm\theta}_j^{(k)} ) = \frac{ \pi_j(\bm r_i; \hat{\bm \tau}^{(k)}) \left[F_{_\text{SMN}}\left(\frac{c_{i2}-\bm x_i^\top \hat{\bm\beta}_j^{(k)} }{\hat{\sigma}_j^{(k)}}; \hat{\bm\nu}_j^{(k)}  \right) - 
				F_{_\text{SMN}}\left(\frac{c_{i1}-\bm x_i^\top \hat{\bm\beta}_j^{(k)} }{\hat{\sigma}_j^{(k)}}; \hat{\bm\nu}_j^{(k)}  \right)  \right]}{ \sum_{l=1}^G \pi_l(\bm r_i;  \hat{\bm\tau}^{(k)}) \left[F_{_\text{SMN}}\left(\frac{c_{i2}- \bm x_i^\top \hat{\bm\beta}_l^{(k)}}{\hat{\sigma}_l^{(k)}};\hat{\bm\nu}_l^{(k)} \right) - F_{_\text{SMN}}\left(\frac{c_{i1}- \bm x_i^\top \hat{\bm\beta}_l^{(k)}}{\hat{\sigma}_l^{(k)}};\hat{\bm\nu}_l^{(k)} \right) \right] }, \\
			\hat{u}_{ij}^{(k)}  & = E(U_{i} |  c_{i1} \leq Y_i \leq c_{i2}, \hat{\bm\theta}_j^{(k)} ), \qquad
			\widehat{uy^2}_{ij}^{(k)} = E(U_{i}Y_i^2|  c_{i1} \leq Y_i \leq c_{i2}, \hat{\bm\theta}_j^{(k)} ), \\
			\widehat{uy}_{ij}^{(k)} & =
			E( U_{i}Y_i |  c_{i1} \leq Y_i \leq c_{i2}, \hat{\bm\theta}_j^{(k)} ),\qquad
			\hat{\Upsilon}_{ij}^{(k)} = E\big(\log h(U_i;\bm\nu_j)|c_{i1} \leq Y_i \leq c_{i2}, \hat{\bm\theta}_j^{(k)} \big).
		\end{align*}
		\end{itemize}
		
		Following \cite{garay2017}, the closed form of the conditional expectations for the particular cases of
		the SMN class of distributions are provided in \ref{appa}.
		
		\item {\bf CM-step 1:} The $M$-step consists of maximizing the $Q$-function with respect to $\bm\Theta^{(k)}$.
		To do this, let $n_j= \sum_{i=1}^n \hat{z}_{ij}^{(k)}$. Then, the maximization of \eqref{qfunc} over $\bm\beta_j$ and  $\sigma_j^2$
		lead to the following CM estimators:
		\begin{align*}
			\hat{\bm\beta}_j^{(k+1)}&= \left( \sum_{i=1}^n \hat{z}_{ij}^{(k)} \hat{u}_{ij}^{(k)} \bm{x}_i \bm{x}_i^\top \right)^{-1}
			\sum_{i=1}^n \hat{z}_{ij}^{(k)}  \widehat{uy}_{ij}^{(k)} \bm{x}_i,\\
			\hat{\sigma}_j^{2(k+1)} &=\frac{1}{n_j} \sum_{i=1}^n \hat{z}_{ij}^{(k)}
			\left(\widehat{uy^2}_{ij}^{(k)} -2 \widehat{uy}_{ij}^{(k)} \bm{x}_i^\top \hat{\bm\beta}_j^{(k+1)} +\hat{u}_{ij}^{(k)}
			\left( \bm{x}_i^\top \hat{\bm\beta}_j^{(k+1)} \right)^2 \right).
		\end{align*}
		\item {\bf CM-step 2:} Following proposition 2 of \cite{Nguyen2016}, the update of $\bm\tau_j$ can be made as
		\[
		\hat{\bm\tau}_j^{(k+1)}= 4\left( \sum_{i=1}^n\bm r_i\bm r_i^\top\right)^{-1}
		\left(\sum_{i=1}^n\left[ \hat{z}_{ij}^{(k+1)}-\pi_j(\bm r_i; \hat{\bm\tau}^{(k)})\right]\bm r_i \right) +\hat{\bm\tau}_j^{(k)}.
		\]
		
		\item {\bf CML-step:} The update of $\bm\nu_j$ crucially depends on the conditional expectation $\hat{\Upsilon}_{ij}^{(k)}$
		which is quite complicated. However, we can update $\bm\upsilon=(\bm\nu_1, \ldots,\bm\nu_G)$ through maximizing
		the actual log-likelihood function as
		\begin{eqnarray}\label{nuupdate}
		\hat{\bm\upsilon}^{(k+1)} &=& \arg\max_{\bm\upsilon} \left\{ \sum_{i=1}^n \log \sum_{j=1}^G \pi_j(\bm r_i;\hat{\bm\tau}^{(k+1)})\left[
		f_{_\text{SMN}}\left(\frac{w_i-\bm x_i^\top\hat{\bm\beta}_j^{(k+1)}}{\hat{\sigma}_j^{(k+1)}}; \bm\nu_j\right)\big/\hat{\sigma}_j^{(k+1)}\right]^{1-\rho_i} \right. \nonumber \\
		&& \qquad \qquad \qquad \left. \left[ F_{_\text{SMN}}\left(\frac{c_{i2}-\bm x_i^\top\hat{\bm\beta}_j^{(k+1)}}{\hat{\sigma}_j^{(k+1)}}; \bm\nu_j\right) - 
		F_{_\text{SMN}}\left(\frac{c_{i1}-\bm x_i^\top\hat{\bm\beta}_j^{(k+1)}}{\hat{\sigma}_j^{(k+1)}}; \bm\nu_j\right)\right]^{\rho_i}\right\}.
		\end{eqnarray}
		Recommended by \cite{Lin2013,Zeller2018}, a more parsimonious model can be achieved by assuming an identical
		mixing component, i.e. $\bm\nu_1=\bm\nu_2=\cdots=\bm\nu_G=\bm\nu$. This setting changes the problem of nontrivial
		high-dimension optimization into the more simple one/two dimension search. The $\texttt{R}$ function $\mathbf{nlminb}(\ )$
		is used to update $\bm\upsilon$ in the numerical parts of the current paper.
	\end{itemize}
	The above E- and M-steps are iterated until some convergence criteria are met. 
	We terminate the algorithm when either the maximum number of iterations approaches l000 or the difference
	between two consecutive log-likelihood values is less than the per-specified tolerance $10^{-5}$.
	
	\begin{remark}\label{rem1}
		To facilitate the update of $\bm\nu=(\nu_1,\ldots,\nu_G)$ for the MoE-CN-CR model in the above EM algorithm, 
		one can introduce an extra latent binary variable $B_{i}$ such that $B_{i}=1$ if an observation $y_i$ in group $g$ is a bad point and 
		$B_{i}=0$ if $y_i$ in group $g$ is a good point. The hierarchical representation of the MoE-CN-CR model can therefore be written as
		\begin{align}\label{CNup}
		Y_i |(\bm{x}_i, U=u_i, {Z}_{ij}=1,B_{i}=1) &\sim \mathcal{N}(\bm x_i^\top\bm\beta_j, u_i^{-1}\sigma_j^2),\nonumber \\
		U_i| (Z_{ij}=1,B_{i}=1) &\sim h(\cdot;\nu_j,\gamma_j),\nonumber \\
		B_{i}| (Z_{ij}=1) &\sim \mathcal{B}(1,\nu_j),\nonumber \\
		Z_i | \bm r_i &\sim \mathcal{M} \left( 1; \pi_1(\bm r_i,\bm\tau),\ldots, \pi_G(\bm r_i,\bm \tau)\right),
		\end{align}
		where $\mathcal{B}(1,\nu_j)$ denotes the Bernoulli distribution with succeed probability $\nu_j$. 
		Consequently, by computing the $Q$-function
		based on \eqref{CNup}, the update of $\nu_j$ is
		\[
		\hat{\nu}_j^{(k+1)} = \frac{\sum_{i=1}^n\hat{z}_{ij}^{(k)}\hat{b}_{ij}^{(k)}}{\sum_{i=1}^n\hat{z}_{ij}^{(k)}},
		\]
		where
		\begin{equation*}
		\hat{b}_{ij}^{(k)} = \left\lbrace \begin{array}{lll}
		\dfrac{\hat{\nu}_j^{(k)} \phi\big(y_i;\bm x_i^\top\hat{\bm\beta}_j^{(k)}, \hat{\gamma}_j^{-1(k)}\hat{\sigma}_j^{2(k)}\big)}{\hat{\nu}_j^{(k)} \phi\big(y_i;\bm x_i^\top\hat{\bm\beta}_j^{(k)}, \hat{\gamma}_j^{-1(k)}\hat{\sigma}_j^{2(k)}\big)+(1-\hat{\nu}_j^{(k)}) \phi\big(y_i;\bm x_i^\top\hat{\bm\beta}_j^{(k)}, \hat{\sigma}_j^{2(k)}\big)},\qquad\qquad \quad
		\text{for the uncensoed cases,}\\
		\\
		\dfrac{\hat{\nu}_j^{(k)} \Big(\Phi\big(c_{i2};\bm x_i^\top\hat{\bm\beta}_j^{(k)},\hat{\gamma}_j^{-1(k)} \hat{\sigma}_j^{2(k)}\big)-\Phi\big(c_{i1};\bm x_i^\top\hat{\bm\beta}_j^{(k)},\hat{\gamma}_j^{-1(k)} \hat{\sigma}_j^{2(k)}\big)\Big)}{F_{CN}\big(c_{i2};\bm x_i^\top\hat{\bm\beta}_j^{(k)}, \hat{\sigma}_j^{2(k)},\hat{\nu}_j^{(k)} ,\hat{\gamma}_j^{(k)}\big)-F_{CN}\big(c_{i1};\bm x_i^\top\hat{\bm\beta}_j^{(k)}, \hat{\sigma}_j^{2(k)},\hat{\nu}_j^{(k)} ,\hat{\gamma}_j^{(k)}\big)},
		\qquad \text{for the censoed cases.}
		\end{array}
		\right. 
		\end{equation*}
		Since there is no closed-form solution for $\hat{\gamma}_j^{(k+1)}$, we update $\gamma_j$ by maximizing the constrained actual 
		observed log-likelihood function \eqref{nuupdate} as a function of $\bm\gamma=(\gamma_1,\ldots,\gamma_G)$.
	\end{remark}

	\subsection{Computational and operational aspects}
	
	\subsubsection{Model selection and performance assessment}
	In practical model-based clustering, it is common to fit a mixture model for the various values of number of
	components $G$ and choose the best $G$ based on some likelihood-based criteria. The two commonly used measures,
	Akaike information criterion (AIC; \cite{akaike1974}) and Bayesian information criterion (BIC; \cite{schwarz1978}),
	are exploited to determine the most plausible value of $G$. The AIC and BIC can be computed as
	\[	\text{AIC} =  2 m - 2\ell_{\max}\qquad \text{BIC} =  m \ln n - 2\ell_{\max},	\]
	where $\ell_{\max}$ is the maximized (observed) log-likelihood, $m$ is the number of free parameters in the model,
	and $n$ is the sample size. Although the smallest value of AIC or BIC results in the most favored model,
	they do not necessarily correspond to optimal clustering.
	For the sake of determining the classification performance, we use the misclassification error rate (MRC), Jaccard 
	coefficient index (JCI; \cite{niwattanakul2013}), Rand index (RI; \cite{rand1971}) and adjusted 
	Rand index (ARI; \cite{hubert1985}) that are computed by comparing predicted classifications to true group labels,
	when known. Noted that the lower MCR (close to zero) or a higher RI and JCI (tend to one) indicates a much similarity 
	between the true labels and the cluster labels obtained by the candidate model.	
	An ARI of one also corresponds to perfect agreement, and the expected value of the ARI under random
	classification is zero. Negative ARI values are possible and indicate classification results that are worse,
	in some sense, than would be expected by random classification.
	
	\subsubsection{Note on computing conditional expectations} 
	As expressed in \ref{appa}, the conditional expectations of the special cases of the
	MoE-SMN-CR model critically depend on the hazard function or the cdf of SMN model. For instance, in the left-censoring 
	scheme, $\widehat{uy}_{ij}^{(k)}$ for the MoE-N-CR model depends on the hazard function of normal distribution as
	${\phi\left(x\right)}\big/{\Phi\left(x\right)}$. The computation of this hazard function for very small values of 
	$x$ (say $x<-35$ as encountered many times in the simulation studies) in $\mathtt{R}$ may lead to ``NaN". To overcome this issue, \cite{Zeller2018} in the $\mathtt{R}$ 
	package ``\textbf{CensMixReg}" set the denominator to the small machine value (the $\mathtt{R}$ commend ``.Machine\$double.xmin"
	was used). However, this setting may lead to negative value for $\hat{\sigma}^2$ as we found. 
	We recommend to use a remedy for obtaining the exact values of 
	$HF(x) = {\phi\left(x\right)}\big/{\Phi\left(x\right)}$. In our computation, we have used log-transformation via
	the following $\mathtt{R}$ command
	\[
	HF(x) = \exp\Big(\text{dnorm}(x, log=T) - \text{pnorm}(x, log.p=T)\Big).
	\]
	
	Figure \ref{HF} in the \ref{appab} highlights the difference of three ways of the HF computation in $\mathtt{R}$. What is
	observed from Figure \ref{HF} is actually the difference between the computation of $HF(x)$ function for $x<-35$. Similar
	trick can be applied for the right- and interval-censoring schemes.  
	
	\section{Monte-Carlo simulation studies}\label{sec4}
		In this section, four Monte-Carlo simulation studies are conducted in order to verify the asymptotic properties of the ML estimates, 
	to assess the fitting and clustering performance of the model, and to check the robustness of the proposed model in dealing with 
	highly peaked, heavily tailed data as well as its sensitivity in presence of outliers.
	
	\subsection{Data generation}
 We note that 
	one of the simplest and straightforward way for generating interval-censored data is to consider 
	$C_{i1} = Y_i - U_i^{(1)}$ and $C_{i2} = Y_i + U_i^{(2)}$ where $U_i^{(1)}$ and $U_i^{(2)}$ are two 
	independent continuous variables followed by $\mathcal{U}(0, c)$ such that the non-informative condition (1.2) of \cite{Gomez2009} 
	is fulfilled. Here $\mathcal{U}(a,b)$ represents the uniform distribution on interval $(a,b)$.
	Recommended by \cite{Gomez2009}, a way to go around non-informative condition is to construct 
	$C_{i1} = \max( Y_i - U_i^{(1)}, Y_i + U_i^{(2)} -c)$ and $C_{i2} = \min(Y_i + U_i^{(2)}, Y_i - U_i^{(1)}+c)$ with $c = 1$,
	which can be shown that fulfills the non-informative condition.
	In short, suppose we generate $n$ realizations from model \eqref{MoE-SMN}, $\bm y=(y_1,\ldots,y_n)^\top$. To have a $p\%$ interval-censored data,
	the following steps are used in our simulation studies. 
\begin{mdframed}[style=MyFrame]
	\begin{itemize}
		\item[$S_1$)] Calculate the number of censored samples $\mathcal{NC}= [n \times p]+1$, where $[a]$ denotes the largest integer 
		not greater than $a$. Then, generate an index set, $\mathcal{IND}$,
		as a sample of size $\mathcal{NC}$ from $\{1, 2, \cdots, n\}$ without replacement. Use $sample(\ )$ function in $\mathtt{R}$ for this 
		purpose. 
		\item[$S_2$)] For $i= 1, \ldots,n$, if $i \in \mathcal{IND}$, then 
		\begin{itemize}
			\item[$S_{21}$)]  Generate two independent random variables, $U_i^1$ and $U_i^2$, from $\mathcal{U}(0,c)$.
			\item[$S_{22}$)] Set 
			$C_{i1} = \max( Y_i - U_i^{(1)}, Y_i + U_i^{(2)} -c), \qquad C_{i2} = \min(Y_i + U_i^{(2)}, Y_i - U_i^{(1)}+c) $.	
		\end{itemize}
	\end{itemize}
\end{mdframed}

	\subsection{Asymptotic properties of the ML estimates}	
	In this section, a simulation study is performed to examine the asymptotic properties of ML parameter
	estimates obtained through the ECME algorithm. We simulate 500 Monte-Carlo samples from the special cases of the MoE-SMN-CR model with $G=2$. 
	The presumed parameters are 
	\[
	\bm\beta_1=(0, -1, -2, -3)^\top,\;\;\bm\beta_2=(-1, 1, 2, 3)^\top, \;\; \sigma_1^2 =1, \;\; \sigma_2^2 =2,\;\; \bm\tau=(0.7, 1, 2)^\top,
	\]
	$\nu_1=\nu_2=3$ for the T and SL distributions, and $(\nu,\gamma)=(0.3,0.3)$ for the CN model.
	For each sample size $n = 50, 100, 500, 2000$, we also set up $\bm x_i = (1, x_{i1}, x_{i2}, x_{i3})^\top$ 
	and $\bm r_i = (1, r_{i1}, r_{i2})^\top$, such that $x_{i1} \sim \mathcal{U}(1,5)$, $x_{i2} \sim \mathcal{U}(-2,2)$, $x_{i3} \sim \mathcal{U}(1,4)$, 
	$r_{i1} \sim \mathcal{U}(-2,1)$ and $r_{i2} \sim \mathcal{U}(-1,1)$. 
	By imposing three levels of right-censoring $(7.5\%, 15\%, 30\% )$ on the data, the ECME algorithm described in Section \ref{ECMsec}
	is preformed to carry out the ML parameter estimates.
	
\begin{figure}\vspace{-0.5cm}
		\begin{center}
			\includegraphics[height=16cm, width=10.5cm, scale=0.5, angle=-90]{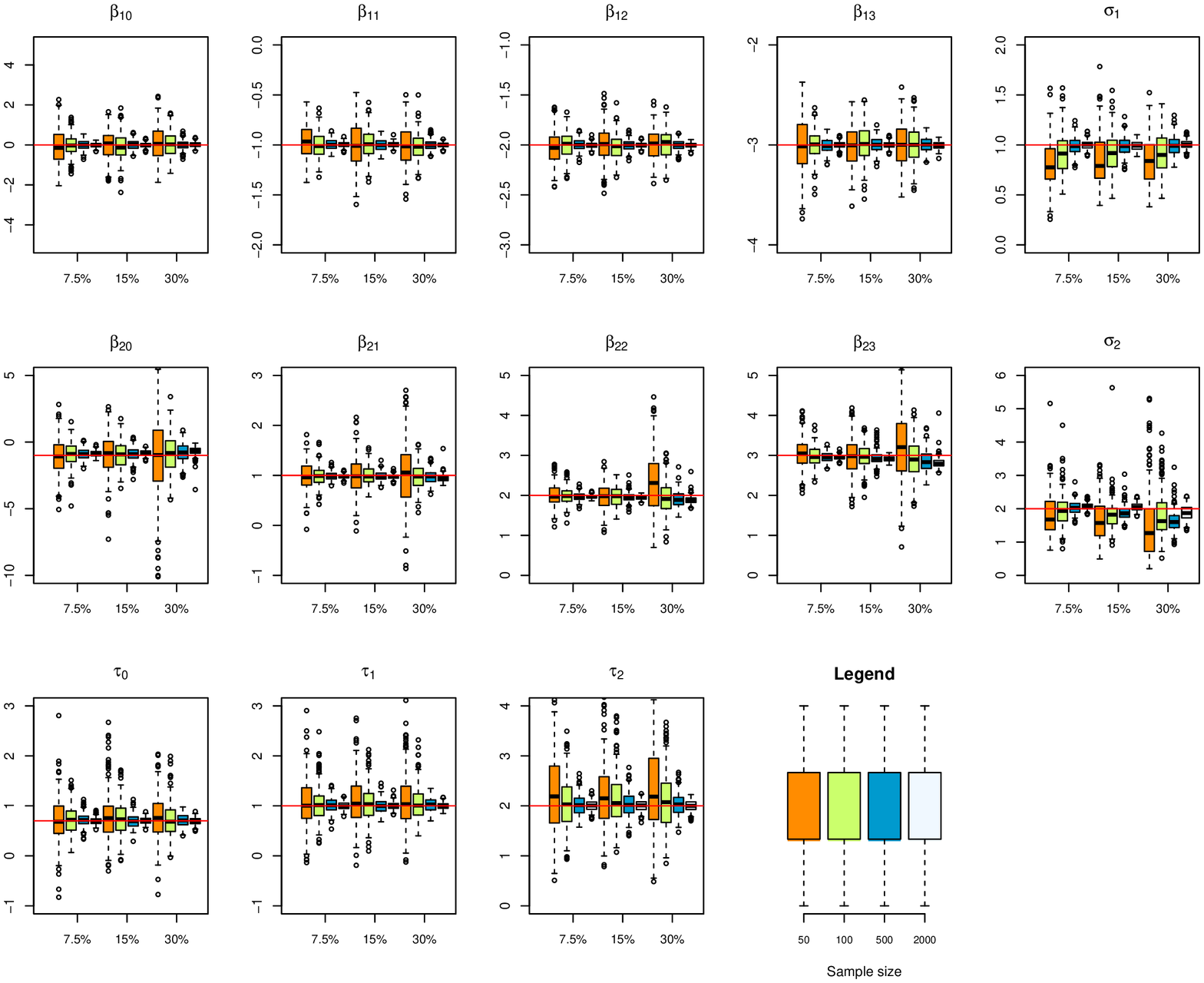}
		\end{center}
		\vspace{-0.5cm}
		\caption{Boxplots of the ML parameter estimates for the MoE-N-CR model.}\label{fig1}
		\begin{center}\vspace{-0.3cm}
			\includegraphics[height=16cm, width=10.5cm, scale=0.5, angle=-90]{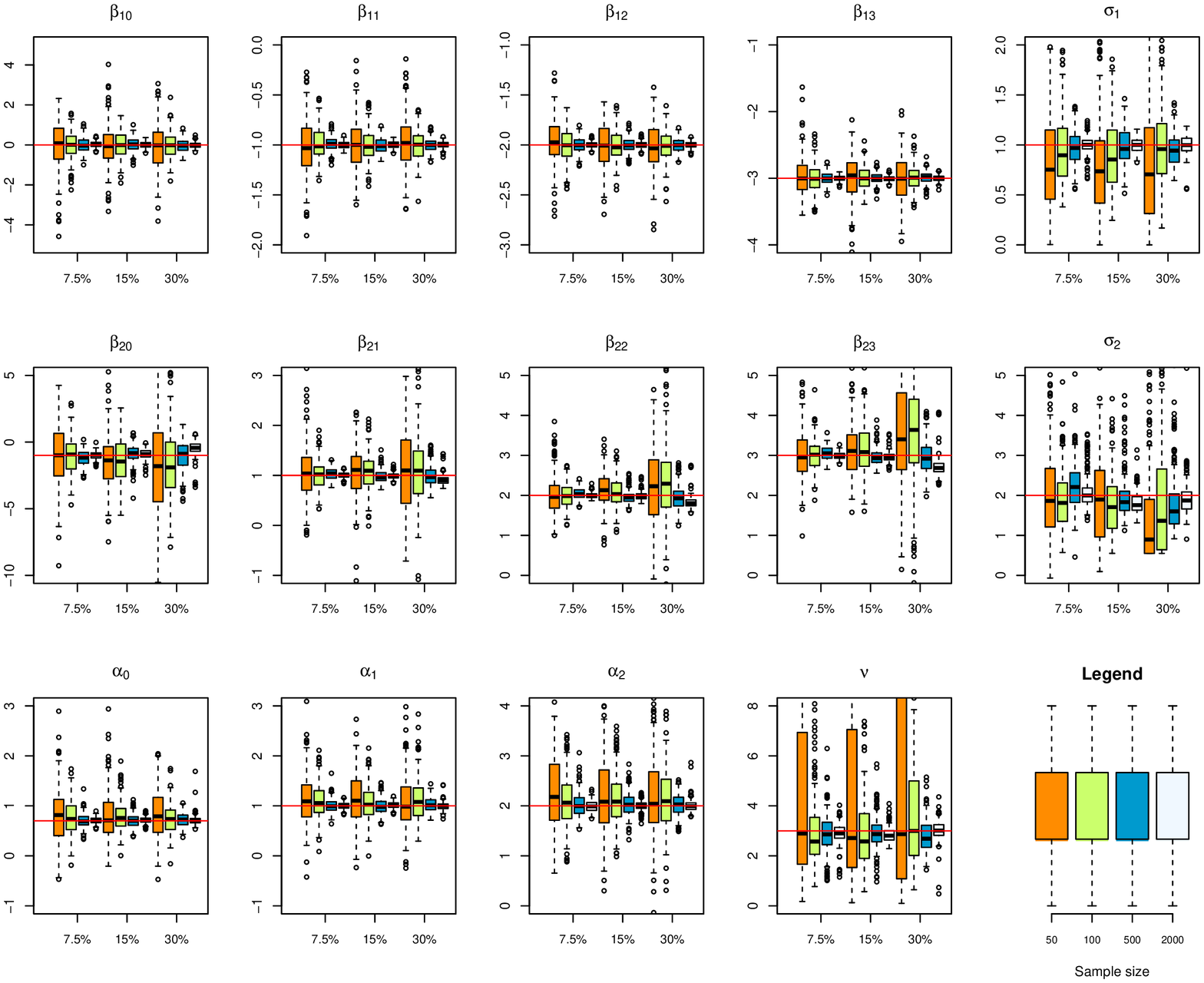}
		\end{center}
		\vspace{-0.5cm}
		\caption{Boxplots of the ML parameter estimates for the MoE-T-CR model.}\label{fig2}
\end{figure}

\begin{figure}
	\begin{center}
		\includegraphics[height=16cm, width=9cm, scale=0.5, angle=-90]{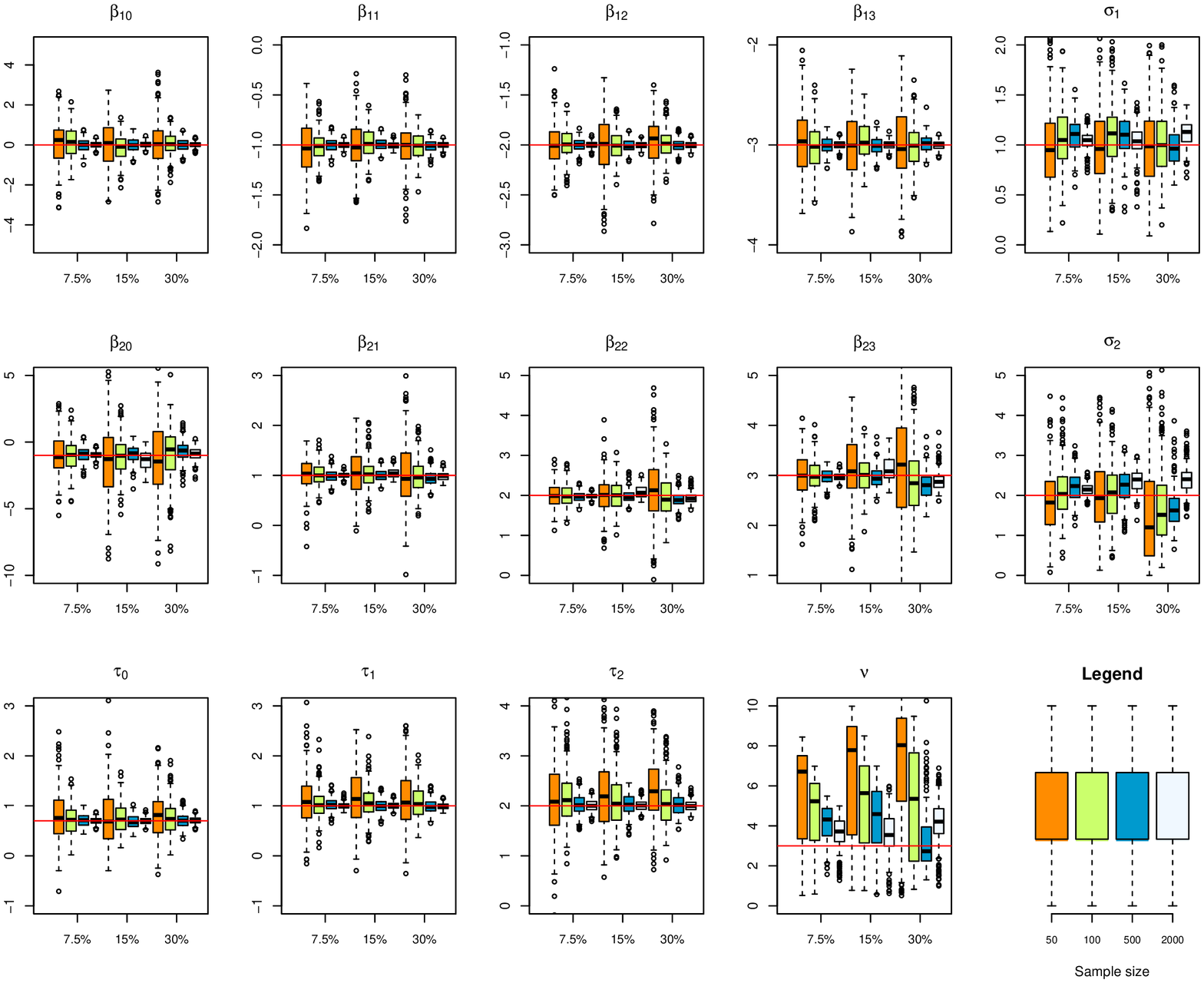} 
	\end{center}
	\vspace{-0.4cm}
	\caption{Boxplots of the ML parameter estimates for the MoE-SL-CR model.}\label{fig3}
	\begin{center}
		\includegraphics[height=16cm, width=9cm, scale=0.5, angle=-90]{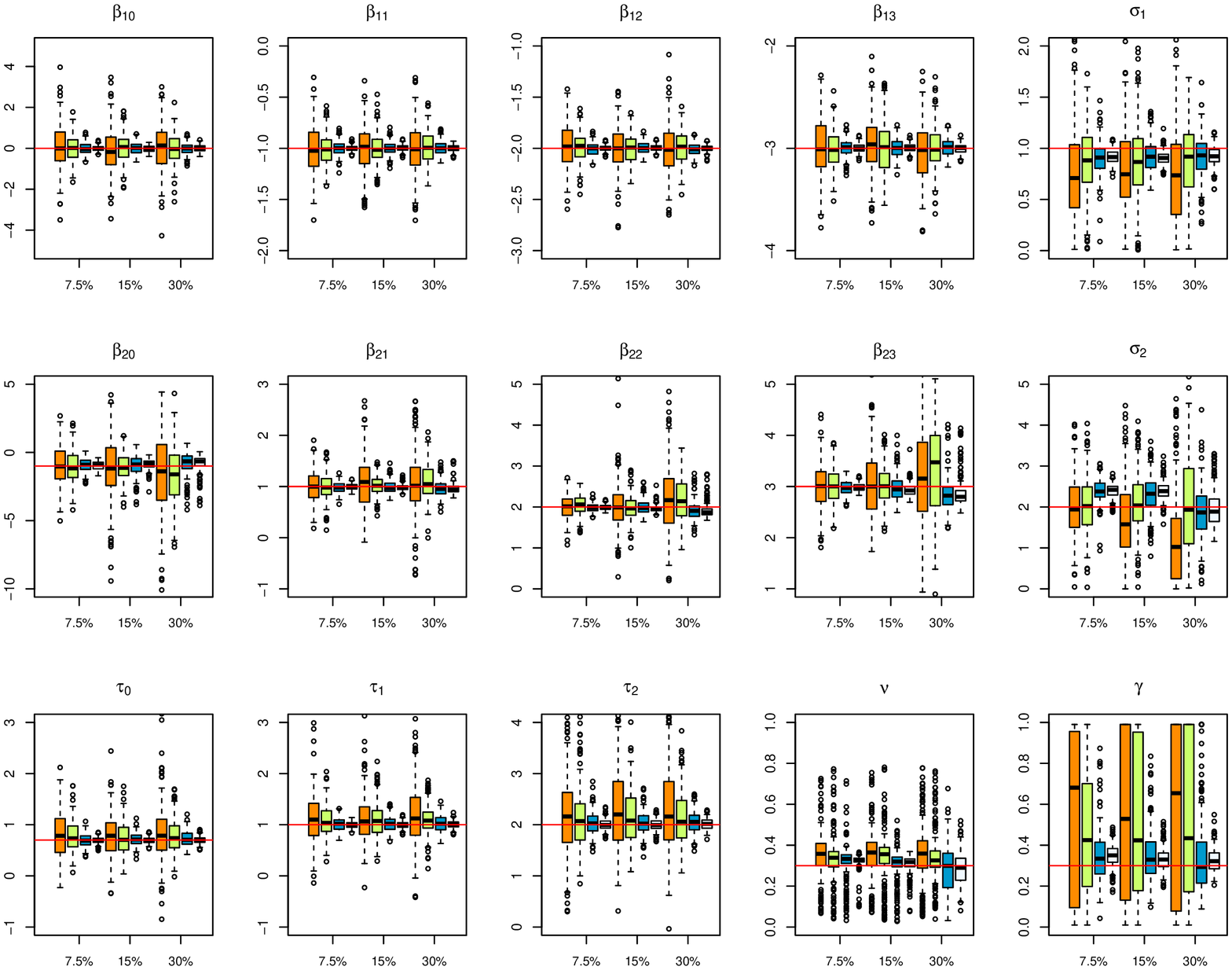}
	\end{center}
	\vspace{-0.4cm}
	\caption{Boxplots of the ML parameter estimates for the MoE-CN-CR model.}\label{fig4}
\end{figure}

	Figures \ref{fig1}-\ref{fig4} display the boxplots of the parameter estimates for the MoE-N-CR, MoE-T-CR, MoE-SL-CR and
	MoE-CN-CR models, respectively. Each plot contains three censoring levels 7.5\%, 15\% and 30\% with four colored boxplots
	representing the sample size of 50 to 2000 from the left to right. It is noticeable that the influence of the censoring in 
	the bias and variability of the parameter estimates increases as the censorship rate increases for all models. 
	As can be expected, the bias and variability tend to decrease toward zero by increasing the sample size, showing empirically 
	the consistency of the ML estimates obtained via the ECME algorithm. It can be also seen that the estimate of the mixing 
	component's parameter for the MoE-T-CR and MoE-SL-CR models has a large bias and variability, especially for small sample sizes.
	Although the estimate of $\gamma$ in the MoE-CN-CR model is bias with large variability as well, it could be argued that the 
	procedure of estimating $\bm\nu$ presented in Remark \ref{rem1} provides a good alternative platform which significantly reduces the 
	bias and variability. 
	

\subsection{Model selection performance via information criteria}\label{sec4.2} 
\begin{figure}[!t]
	\begin{center}
		\includegraphics[height=6cm, width=7cm]{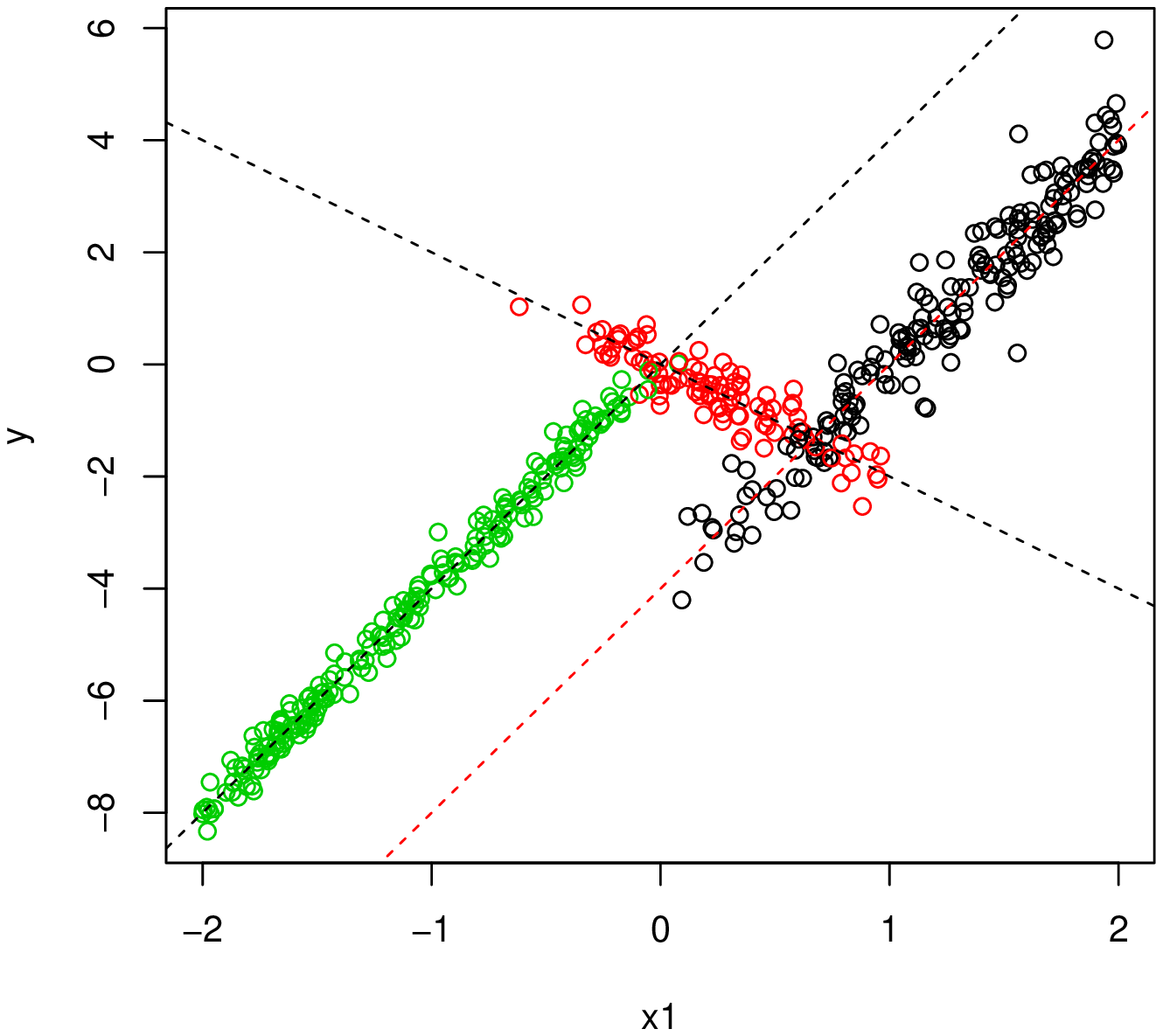}
		\includegraphics[height=6cm, width=7cm]{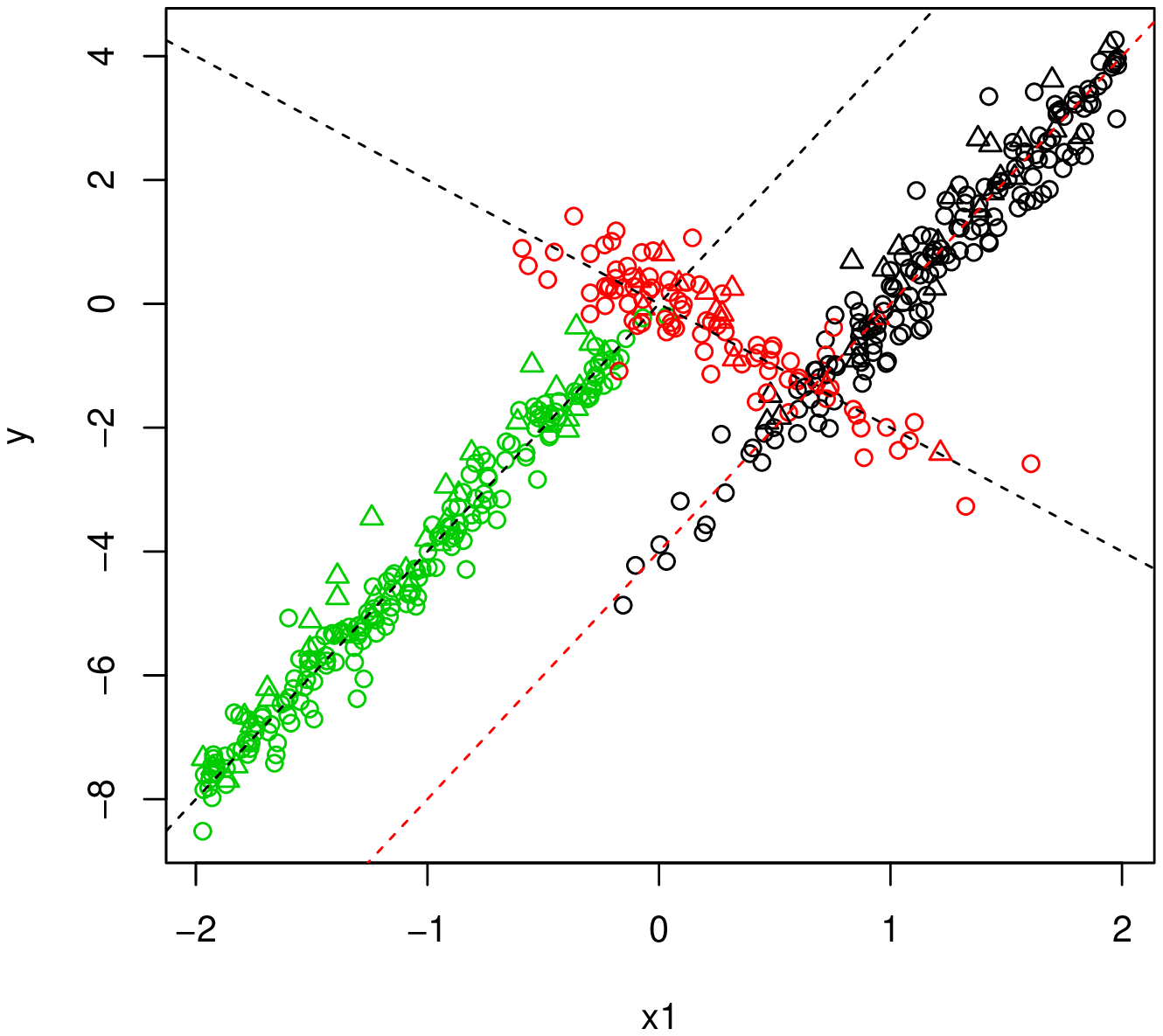} 
	\end{center}
	\vspace{-0.7cm}
	\caption{Simulated MoE-SGIG-CR data. Left-top panel: data without any censored observation.
		Right-top panel: data with 15\% left-censored observations denoted by $\bigtriangleup$. Dash lines represent the
		true experts.}\label{fig5}
\end{figure}

One of the challenges in the MoE models is to choose the optimal number of experts $G$.
In dealing with this challenge, we conduct a simulation study to compare the ability of the proposed sub-models in the 
class of MoE-SMN-CR model to select the accurate $G$. We generate 100 sample of size $n=500$ from a three components (i.e. true $G$=3)
MoE-SMN-CR model \eqref{MoE-SMN}, where the mixing variable $U$ is followed by a generalized inverse Gaussian 
(GIG) distribution with parameters $\theta=(\kappa,\chi,\psi)$, denoted by the MoE-SGIG-CR model. Details of 
the GIG distribution and its new data-generating algorithm can be found in \cite{Hoermann2013}. It is assumed 
that the data is left-censored with levels 7.5\%, 15\% or 30\%, $\bm x_i = \bm r_i = (1, x_{i1})^\top$ 
such that $x_{i1}\sim \mathcal{U}(-2, 2)$, $\bm\beta_1=(-4, 4)^\top,$ $\bm\beta_2=(0,-2)^\top$, $\bm\beta_3=(0,4)^\top$, 
$\bm\tau_1=(0, 13)^\top$, $\bm\tau_2=(2,9)^\top$, and $\theta_1=(-0.5,1,2)$, $\theta_2=(0.5,1,2)$, $\theta_3=(-0.5,2,1)$. Example of
generated samples with and without censored cases are shown in Figure \ref{fig5}.

\begin{figure}[!t]
	\begin{center}
		\includegraphics[height=6cm, width=7cm]{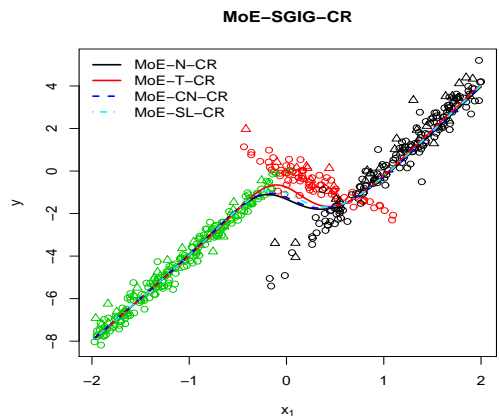}
	\end{center}
	\vspace{-0.7cm}
	\caption{The estimated Experts curves of the special cases of the MoE-SMN-CR model for MoE-SGIG-CR simulated data with
		15\% left-censoring.}\label{fig51}
\end{figure}

In this simulation study, we assume that the number of mixture components $G$ is unknown. 
We therefore fit the MoE-N-CR, MoE-T-CR, MoE-SL-CR and MoE-CN-CR models to the generated data in each replication, by assuming 
$G$ ranging form 1 to 5. The detailed numerical results including the average values of AIC and BIC together with the 
rate of correct model specification (RC; the mean of the number of replications in which the model with $G=3$ is outperformed) 
are reported in Table \ref{tab11}. Based on the RC measure, the MoE-T-CR, MoE-SL-CR and MoE-CN-CR models perform better than the MoE-N-CR model
in identifying the number of components since the data are generated from a heavy-tailed distribution. 
Results depicted in Table \ref{tab11} suggest that 
the BIC is more reliable than the AIC for model selection purpose and based on this measure the MoE-T-CR and MoE-SL-CR models 
outperform the other MoE models to fit to the data. In Figure \ref{fig51}, we plot the curve of the 
estimated experts to a dataset, with 15\% censoring level, in which all models suggest $G=3$ based on the BIC. 
It could clearly be observed that the MoE-T-CR model fit the data better than the other models.

\begin{table*}[!t]
	\centering{\caption{The average of AIC and BIC, over 100 replications, by fitting special cases of 
			the MoE-SMN-CR model to the generated data from the MoE-SGIG-CR model.\label{tab11}}
		\scalebox{0.65}{\hspace{-0.5cm}
			\begin{tabular}{lcccccccccccccccccccccccccccccccc}
				\hline
				 && \multicolumn{2}{c}{$G=1$} &&
				\multicolumn{2}{c}{$G=2$} && \multicolumn{2}{c}{$G=3$} && \multicolumn{2}{c}{$G=4$} && \multicolumn{2}{c}{$G=5$} && \multicolumn{2}{c}{RC}\\
				\cline{3-4}\cline{6-7}\cline{9-10}\cline{12-13}\cline{15-16}\cline{18-19}
				Model & Cens. Level & AIC & BIC && AIC & BIC && AIC & BIC && AIC & BIC && AIC & BIC&& AIC & BIC\\
				\hline
				MoE-N-CR & $7.5\%$ & 1959.914 & 1972.558 && 1199.770 & 1233.487 && 937.685 & $\bm{992.475}$ && $\bm{918.283}$ & 994.146 && 929.686 & 1026.622 && 0.39 & 0.73\\
				         & $15\%$ & 1950.180 & 1962.310 &&  1201.584 & 1231.051 && 954.333 & $\bm{1007.388}$ && 939.155 & 1011.253 && $\bm{931.957}$ & 1027.453 && 0.35 & 0.59\\
				         & $30\%$ & 1947.912 & 1960.556 && 1164.255 & 1197.972 && 886.857 & $\bm{941.647}$ && 879.832 & 955.695 && $\bm{867.216}$ & 964.152 && 0.35 & 0.78\\
				         \\
				MoE-T-CR & $7.5\%$ & 1944.642 & 1961.501 && 1031.442 & 1073.588 && 915.670 & $\bm{983.104}$ && $\bm{882.016}$ & 974.737 && 893.425 & 1011.434 && 0.58 & 0.83\\
				         & $15\%$ & 1936.854 & 1953.712 && 1037.712 & 1079.858 && $\bm{892.430}$ & $\bm{959.864}$ && 893.762 & 986.483 && 897.911 & 1015.920 && 0.59 & 0.80\\
				         & $30\%$ & 1924.194 & 1941.052 && 1018.596 & 1060.742 && 823.155 & $\bm{890.588}$ && 814.159 & 906.880 && $\bm{810.957}$ & 928.966 && 0.49 & 0.85\\
				         \\
				MoE-SL-CR& $7.5\%$ & 1945.130 & 1961.988 && 1048.648 & 1090.795 && 895.996 & $\bm{963.430}$ && $\bm{883.740}$ & 976.462 && 903.997 & 1022.006 && 0.61 & 0.84\\
			             & $15\%$ & 1937.009 & 1953.867 && 1076.567 & 1118.713 && 908.995 & $\bm{976.428}$ && $\bm{905.568}$ & 998.289 && 911.626 & 1029.635 && 0.62 & 0.78\\
				         & $30\%$ & 1924.250 & 1941.109 && 1044.162 & 1086.308 && 838.302 & $\bm{905.735}$ && $\bm{833.820}$ & 926.542 && 840.034 & 958.043 && 0.55 & 0.85\\
				         \\
				MoE-CN-CR& $7.5\%$ & 1946.769 & 1967.842 && 1094.680 & 1145.256 && 931.721 & $\bm{1011.798}$ && $\bm{910.684}$ & 1020.264 && 913.511 & 1052.592 && 0.58 & 0.77\\
			             & $15\%$ &  1938.737 & 1959.810 && 1136.442 & 1187.017 && 962.042 & $\bm{1042.120}$ && 938.145 & 1047.725 && $\bm{918.808}$ & 1057.890 && 0.40 & 0.70\\
			 	         & $30\%$ & 1927.552 & 1948.625 && 1089.598 & 1140.173 && 876.187 & $\bm{956.264}$ && 871.465 & 981.045 && $\bm{866.797}$ & 1005.879 && 0.54 & 0.90 \\		   
				\hline
			\end{tabular}
	}}
\end{table*}

\subsection{Model performance in dealing with the highly peaked and thick-tailed data}
In this simulation study, we simulate data with $n=100, 500$ and 2000 observations from a three-component MoE-SMN-CR model
via representation \eqref{SMN-lin-rep} under two generating scenarios of $U$. The first scenario (S1) is conducted by assuming 
$U^{-1}\sim \mathcal{E}(0.5)$, the exponential distribution with parameter $\lambda=0.5$, whereas the second one (S2) considers 
$U\sim \mathcal{BS}(\alpha,1)$, the Birnbaum-Saunders distribution \citep{birnbaum1969new} with parameter $\alpha$ and $\beta=1$.
Bear in mind that the former scenario generates 
data from a Laplace distribution which is known as a highly peaked model
and the later scenario provides a heavier tailed model than the normal distribution \citep{Naderi2017}. The Laplace and BS censored
MoE models, referred as the MoE-SLap-CR and MoE-SBS-CR, are not considered in this paper since their conditional expectations involved 
in the ECME algorithm are not exist.  

In each replication of 200 trials, we generate data from the MoE-SLap-CR and MoE-SBS-CR models with three components, 
interval-censoring levels 7.5\%, 15\% or 30\%, 	$\bm x_i = (1, x_{i1}, x_{i2},x_{i3})^\top$ 
and $\bm r_i = (1, r_{i1})^\top$ , 
such that $x_{i1} \sim \mathcal{U}(1,5)$, $x_{i2} \sim U(0,1)$, $x_{i3} \sim U(-2,-1)$,
and $r_{i1} \sim \mathcal{U}(-1,1)$, presumed parameter values given by
$\bm\theta_j=(\bm\beta_j,\sigma^2_j,\bm\nu_j),~j= 1,2,3$, $\bm\beta_1= (-2, -1, -2, -3)^\top$,  
$\bm\beta_2 = (0.5, 1, 2, 3)^\top$, $\bm\beta_3 = (2, 1, 3, 5)^\top$, $\sigma^2_1 = 1$, $\sigma^2_2 = 3$,
$\sigma^2_3 = 5$, $\bm\tau_1= (2, 10)^\top$, $\bm\tau_2= (0.7, 10)^\top$ and $(\alpha_1,\alpha_2,\alpha_3)=(3,1,2)$ 
for the MoE-SBS-CR model.	

We compare the performance of the three-component MoE-N-CR, MoE-T-CR, MoE-SL-CR, and MoE-CN-CR models in terms of model selection
indices (AIC and BIC) as well as clustering agreement measures (MCR, JCI, and ARI). Tables \ref{tab1} and \ref{tab2} present the
average values of AIC, BIC, MCR, JCI and ARI over all 200 replications for the S1 and S2 scenarios of simulation, respectively.
Results depicted in these tables reveal that the MoE-T-CR model outperforms the others in terms of AIC and BIC.
Although the clustering performance of all models are very closed to each others, as expected from the MoE structurer, 
the MoE-T-CR and MoE-CN-CR models provide a slight improvement in the MCR, JCI and AIR over the MoE-N-CR and MoE-SL-CR models.

\begin{table*}[!t]
	\centering{\caption{The average of AIC, BIC, MCR and AIR, over 200 replications, by fitting special cases of 
			the MoE-SMN-CR model to the generated data under S1 scenario.\label{tab1}}
		\scalebox{0.65}{\hspace{-0.5cm}
			\begin{tabular}{cccccccccccccccccccccccccccccccccc}
				\hline
				Model  $\rightarrow$& & \multicolumn{3}{c}{MoE-N-CR} & &
				\multicolumn{3}{c}{MoE-T-CR} & & \multicolumn{3}{c}{MoE-SL-CR} & & \multicolumn{3}{c}{MoE-CN-CR}\\
				\cline{3-5}\cline{7-9}\cline{11-13}\cline{15-17}
				$n\;\downarrow$ & Measure & $7.5\%$ & $15\%$ & $30\%$ && $7.5\%$ & $15\%$ & $30\%$
				&& $7.5\%$ & $15\%$ & $30\%$ && $7.5\%$ & $15\%$ & $30\%$ \\
				\hline
				& AIC & 496.030 & 505.090 & 514.104 && 485.298 & 496.103 & 497.694 && 494.472 & 503.898 & 506.189 && 502.770 & 511.464 & 514.687\\
				& BIC & 545.528 & 554.588 & 563.602 && 542.612 & 553.417 & 555.008 && 551.786 & 561.212 & 563.503 && 567.901 & 576.594 & 579.817\\
				100 & MCR & 0.165 & 0.192 & 0.222 &&  0.175 & 0.185 & 0.225 && 0.165 & 0.178 & 0.216 && 0.158 & 0.175 & 0.213\\ 
				& ARI & 0.618 & 0.576 & 0.499 && 0.598 & 0.584 & 0.497 && 0.617 & 0.594 & 0.506 && 0.630 & 0.604 & 0.517\\
				& JCI & 0.632 & 0.602 & 0.551 && 0.619 & 0.608 & 0.548 && 0.630 & 0.616 & 0.553 && 0.641 & 0.621 & 0.560\\
				\\
				& AIC & 2370.929 & 2425.217 & 2618.009 && 2330.694 & 2369.694 & 2531.162 && 2338.787 & 2383.382 & 2554.371 && 2335.274 & 2378.907 & 2545.153\\
				& BIC & 2451.006 & 2505.295 & 2698.086 && 2423.415 & 2462.416 & 2623.884 && 2431.509 & 2476.104 & 2647.092 && 2440.640 & 2484.272 & 2650.518\\
				500 & MCR & 0.162 & 0.167 & 0.214 && 0.153 & 0.163 & 0.201 && 0.154 & 0.158 & 0.186 && 0.155 & 0.157 & 0.187 \\
				& ARI & 0.614 & 0.616 & 0.572 && 0.628 & 0.617 & 0.577 && 0.627 & 0.624 & 0.598 && 0.626 & 0.629 & 0.596 \\
				& JCI & 0.630 & 0.634 & 0.597 && 0.642 & 0.635 & 0.601 && 0.642 & 0.641 & 0.616 && 0.639 & 0.642 & 0.612 \\	
				\\
				& AIC & 9397.490 & 9559.804 & 10548.390 && 9220.558 & 9278.635 & 10085.960 && 9255.169 & 9325.645 & 10221.340 && 9235.637 & 9310.615 & 10157.243\\
				& BIC & 9503.907 & 9666.221 & 10654.800 && 9343.778 & 9401.854 & 10209.180 && 9378.389 & 9448.865 & 10344.560 && 9370.060 & 9445.036 & 10291.67\\  
				2000& MCR & 0.154 & 0.167 & 0.240 && 0.147 & 0.159 & 0.222 && 0.147 & 0.159 & 0.213 && 0.146 & 0.154 & 0.202\\  
				& ARI & 0.634 & 0.614 & 0.521 && 0.644 & 0.621 & 0.530 && 0.645 & 0.623 & 0.541 && 0.647 & 0.623 & .556\\
				& JCI & 0.645 & .0632 & 0.559 && 0.654 & .0637 & 0.564 && 0.656 & 0.641 & 0.571 && 0.656 & 0.644 & 0.580 \\	
				\hline
			\end{tabular}
	}}
\end{table*}

\begin{table*}[!t]
	\centering{\caption{The average of AIC, BIC, MCR and AIR, over 200 replications, by fitting special cases of 
			the MoE-SMN-CR model to the generated data under S2 scenario.\label{tab2}}
		
		\scalebox{0.64}{\hspace{-0.75cm}
			\begin{tabular}{cccccccccccccccccccccccccccccccccc}
				\hline
				Model  $\rightarrow$& & \multicolumn{3}{c}{MoE-N-CR} & &
				\multicolumn{3}{c}{MoE-T-CR} & & \multicolumn{3}{c}{MoE-SL-CR} & & \multicolumn{3}{c}{MoE-CN-CR}\\
				\cline{3-5}\cline{7-9}\cline{11-13}\cline{15-17}
				$n\;\downarrow$ & Measure & $7.5\%$ & $15\%$ & $30\%$ && $7.5\%$ & $15\%$ & $30\%$
				&& $7.5\%$ & $15\%$ & $30\%$ && $7.5\%$ & $15\%$ & $30\%$ \\
				\hline
				& AIC & 522.443 & 542.212 & 556.691 && 498.602 & 514.384 & 533.806 && 508.967 & 525.134 & 544.353 && 502.185 & 521.979 & 538.774\\  
				& BIC & 571.941 & 591.710 & 606.189 && 555.916 & 571.698 & 591.121 && 566.280 & 582.448 & 601.666 && 567.315 & 587.108 & 603.903\\  
				100 & MCR & 0.235 & 0.248 & 0.263 && 0.183 & 0.190 & 0.209 && 0.188 & 0.193 & 0.211 && 0.190 & 0.197 &  0.214 \\ 
				& ARI & 0.498 & 0.472 & 0.448 && 0.589 & 0.571 & 0.541 && 0.584 & 0.563 & 0.539 && 0.581 & 0.557 & 0.535\\
				& JCI & 0.518 & 0.501 & 0.493 && 0.627 & 0.599 & 0.572 && 0.618 & 0.586 & 0.577 && 0.605 & 0.572 & 0.561\\
				\\ 
				& AIC & 2561.848 & 2564.510 & 2665.281 && 2432.404 & 2432.566 & 2503.707 && 2455.665 & 2464.471 & 2551.097 && 2490.907 & 2502.043 & 2799.294 \\
				& BIC & 2641.925 & 2644.587 & 2745.358 && 2525.125 & 2525.288 & 2596.428 && 2548.386 & 2557.192 & 2643.818 && 2596.273 & 2607.408 & 2904.659\\
				500 & MCR & 0.197 & 0.207 & 0.217 && 0.162 & 0.172 & 0.187 && 0.172 & 0.179 & 0.190 && 0.176 & 0.186 & 0.201\\
				& ARI & 0.588 & 0.553 & 0.533 && 0.636 & 0.602 & 0.574 && 0.624 & 0.585 & 0.563 && 0.618 & 0.574 & 0.545\\
				& JCI & 0.604 & 0.583 & 0.577 && 0.643 & 0.622 & 0.601 && 0.633 & 0.612 & 0.587 && 0.627 & 0.596 & 0.569\\	
				\\
				& AIC & 10056.300 & 10424.278 & 10954.970 && 9576.709 & 9829.808 & 10326.340 && 9646.046 & 9956.371 & 10514.060 && 9654.003 & 9976.371 & 10527.060 \\
				& BIC & 10162.717 & 10530.695 & 11061.380 && 9699.929 & 9953.028 & 10449.560 && 9769.266 & 10079.591 & 10637.280 && 9794.025 & 10110.791 & 10661.420\\
				2000& MCR & 0.214 & 0.223 & 0.255 && 0.171 & 0.174 & 0.216 && 0.189 & 0.178 & 0.218 && 0.171 & 0.181 & 0.219\\
				& ARI & 0.556 & 0.535 & 0.513 && 0.613 & 0.601 & 0.569 && 0.592 & 0.576 & 0.533 && 0.601 & 0.583 & 0.550 \\
				& JCI & 0.578 & 0.560 & 0.553 && 0.644 & 0.619 & 0.598 && 0.607 & 0.596 & 0.563 && 0.626 & 0.603 & 0.569 \\	
				\hline
			\end{tabular}   
	}}
\end{table*}

\subsection{Sensitivity analysis in presence of outliers}
\begin{figure}[!t]
	\begin{center}
		\includegraphics[height=5cm, width=5cm]{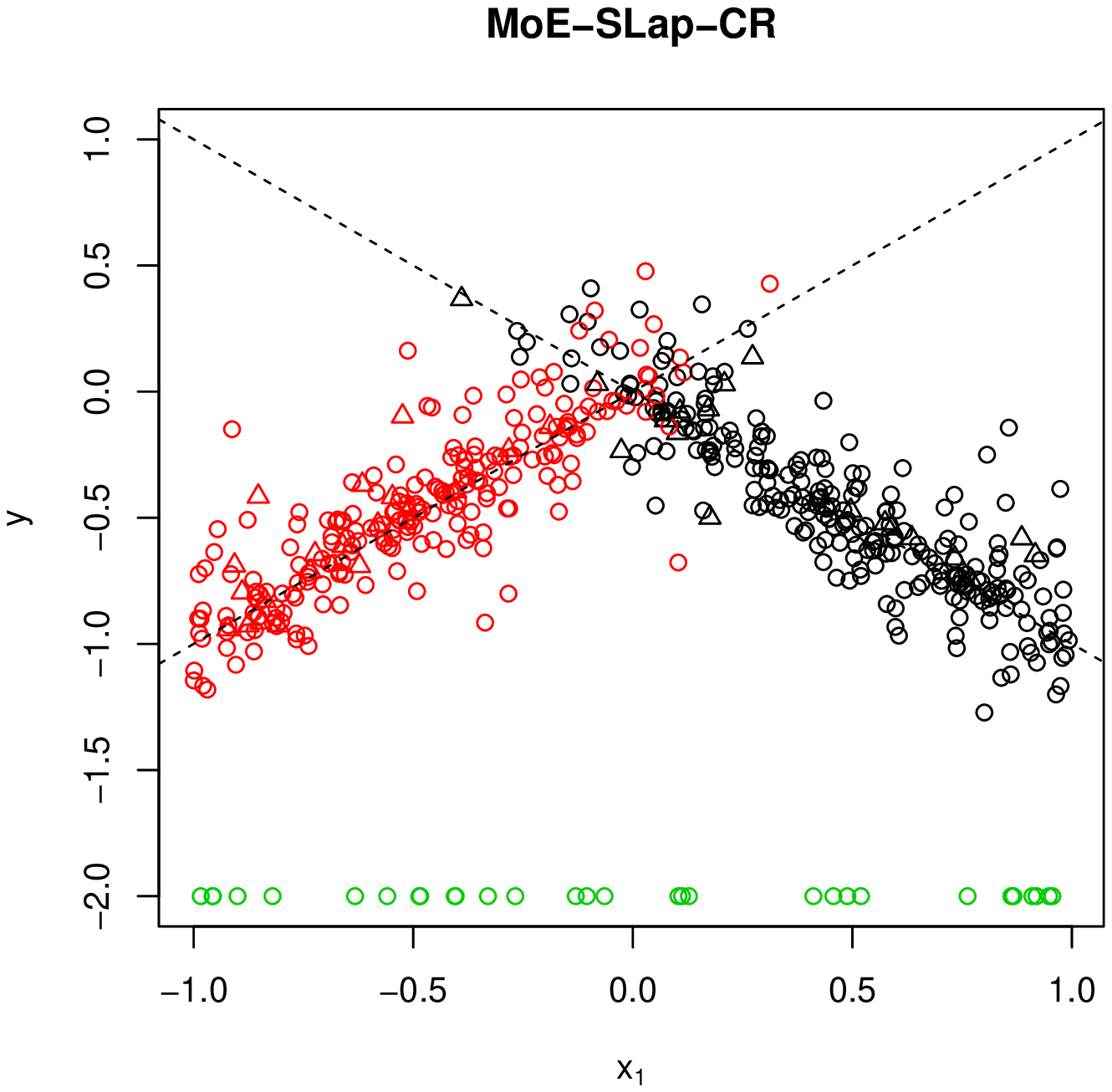} 
		\includegraphics[height=5cm, width=5cm]{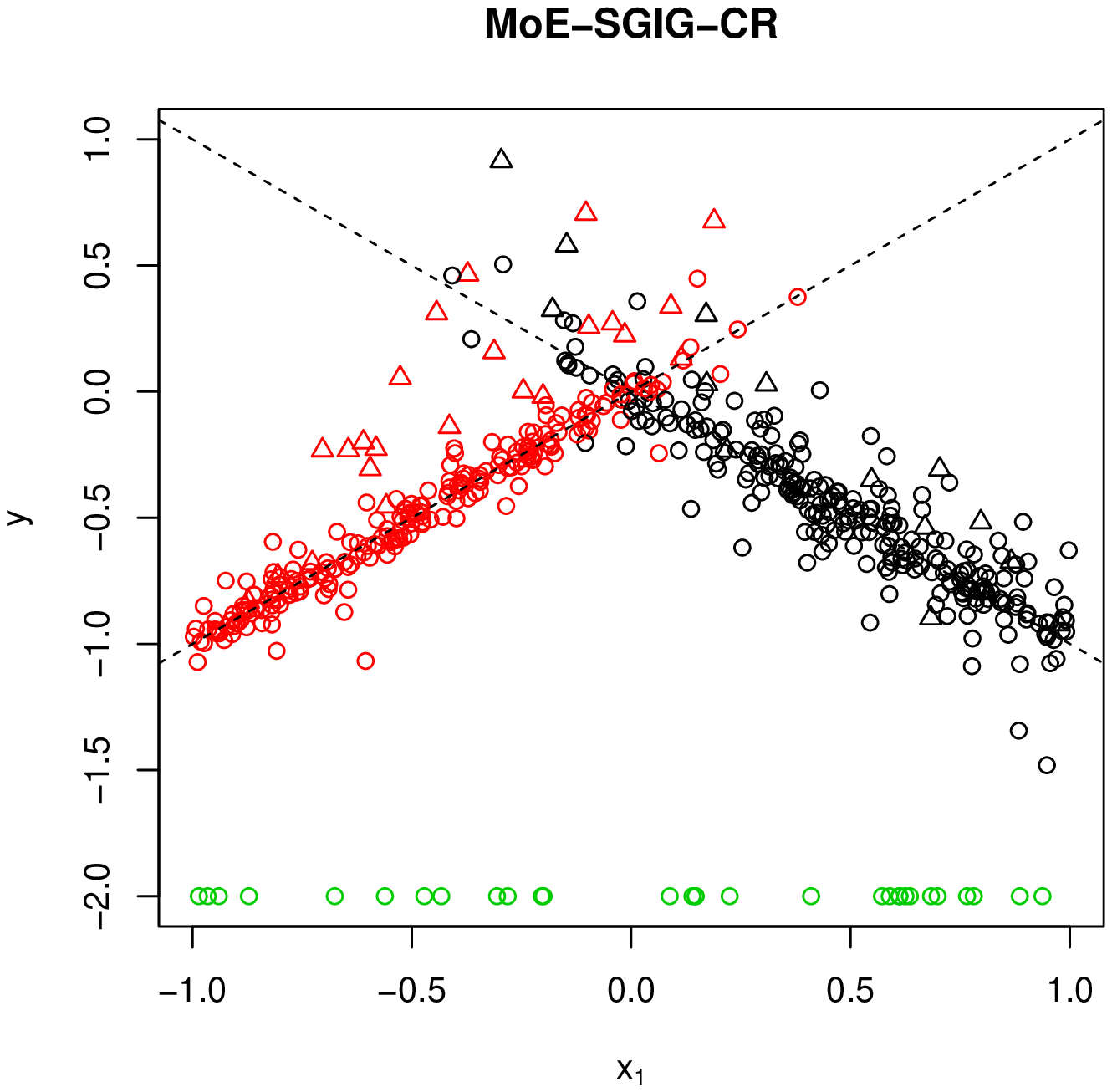}
		\includegraphics[height=5cm, width=5cm]{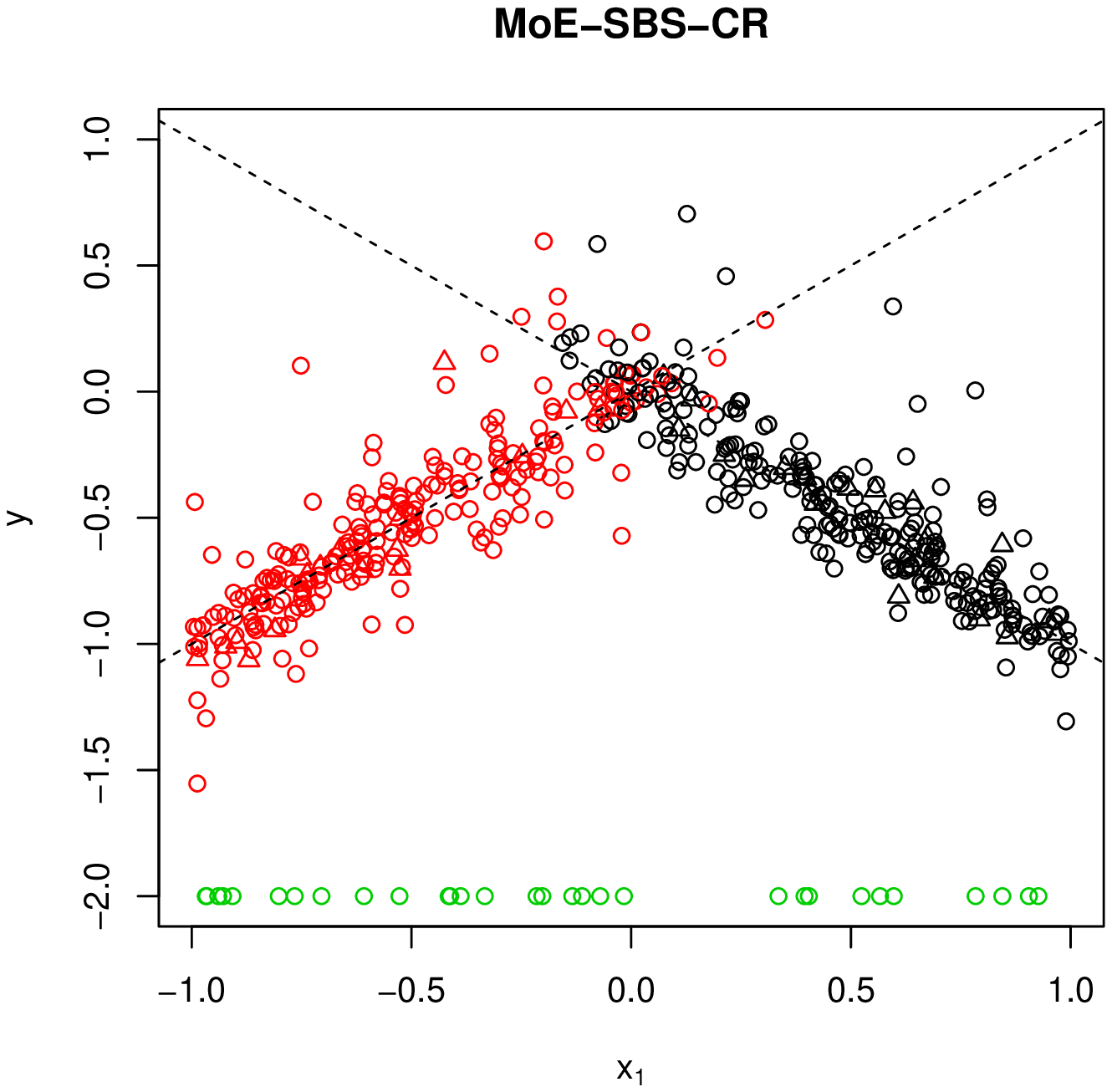} 
	\end{center}
	\vspace{-0.7cm}
	\caption{Scatterplots of the simulated data with 7.5\% left-censoring ($\triangle$) generated from 
		the MoE-SLap-CR, MoE-SBS-CR and MoE-SGIG-CR models and containing 6\% outliers (green $\circ$). Dash lines represent the
	true experts.}\label{Sim4}%
\end{figure}

\begin{figure}[!t]
	\begin{center}
		\includegraphics[height=6cm, width=7cm]{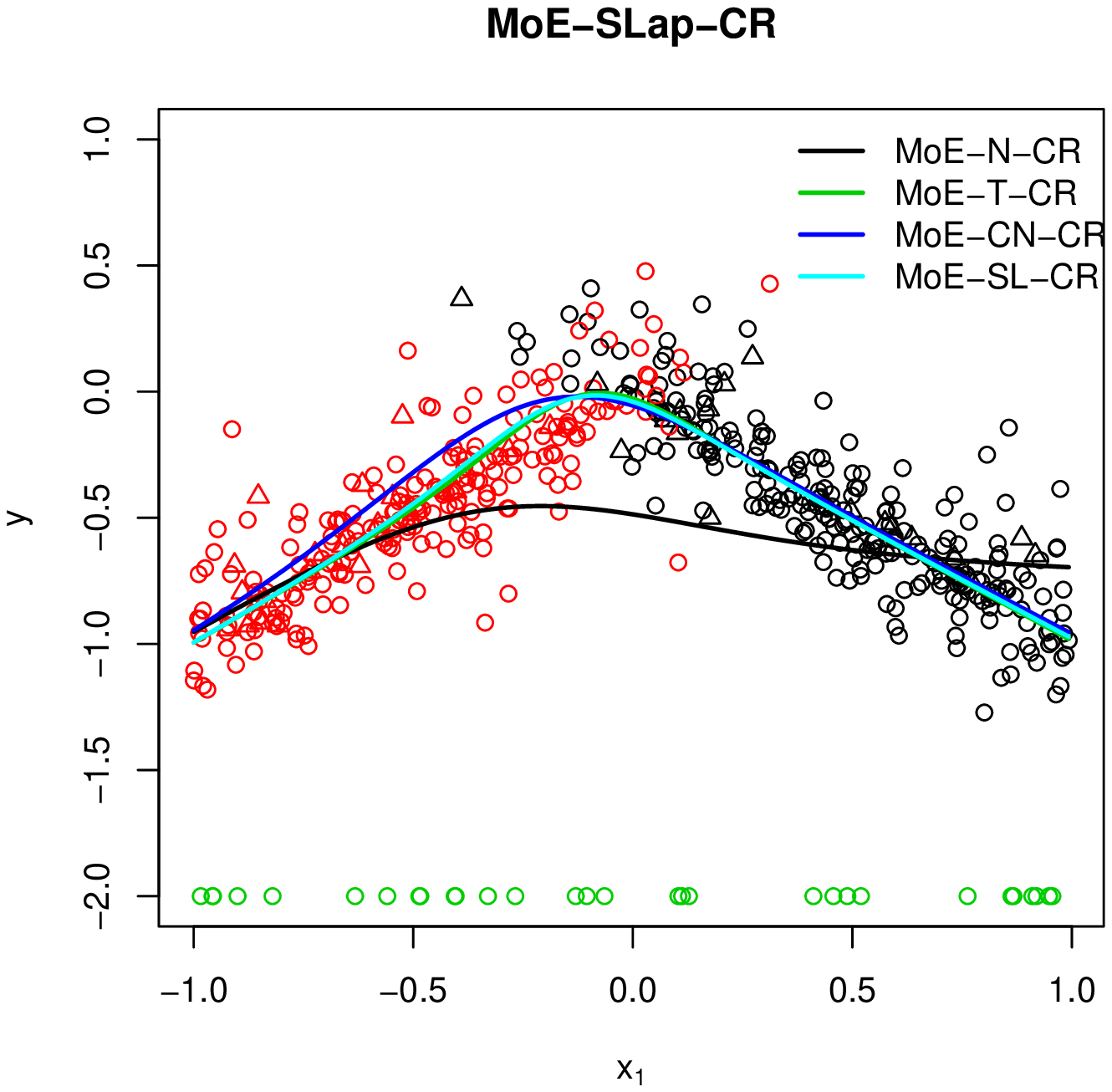} 
		\includegraphics[height=6cm, width=7cm]{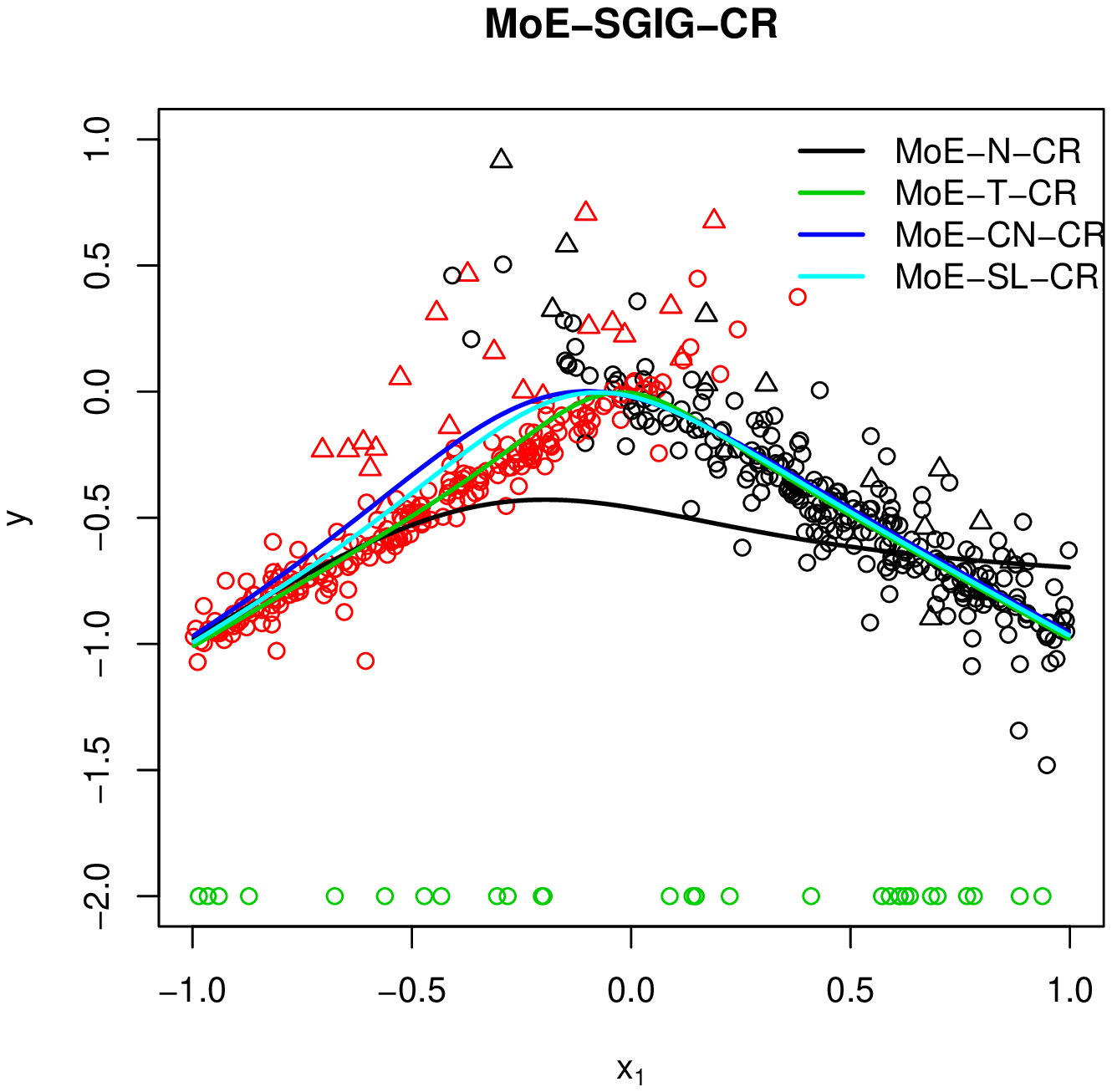}
		\includegraphics[height=6cm, width=7cm]{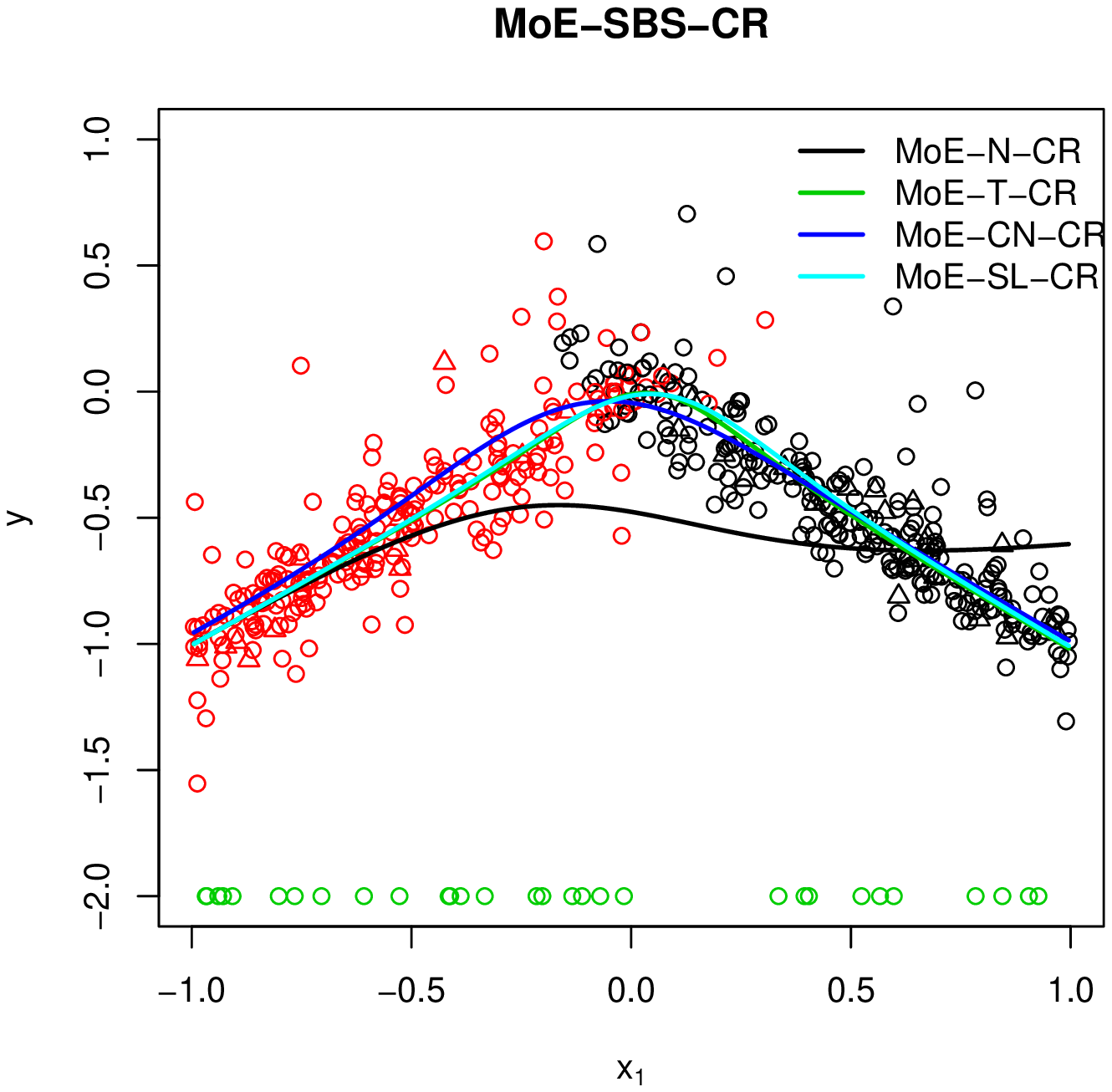} 
	\end{center}
	\vspace{-0.7cm}
	\caption{Scatter plots of the artificial data with 7.5\% left-censoring ($\triangle$) generated from 
		the MoE-SLap-CR, MoE-SBS-CR and MoE-SGIG-CR models and containing 6\% outliers (green $\circ$).}\label{Sim42} %
\end{figure}

The last simulation study aims at investigating the robustness on estimating MoE-SMN-CR models in which some outliers 
are introduced into the simulated data.	Each of the three models MoE-SLap-CR, MoE-SBS-CR and MoE-SGIG-CR is considered
for data generation. Following \cite{Nguyen2016}, we setup $\bm x_i = \bm r_i=(1, x_{i1})^\top$ where $x_{i1}$ is 
generated from $\mathcal{U}(-1,1)$, $\bm\beta_1=(0, 1)^\top,$ $\bm\beta_2=(0,-1)^\top$, $\sigma^2_1=\sigma^2_2=0.01$
$\bm\tau_1=(0, 10)^\top$, $\theta_1=(-0.5,1,0.2)$, $\theta_2=(0.5,1,0.2)$ for the MoE-SGIG-CR model and $(\alpha_1,\alpha_2)=
(0.5,1)$ for the MoE-SBS-CR model. We assume left-censoring scheme with levels  7.5\% or 30\% and sample size 500. 
We also add class of outliers with varying probability $c$ ranging from 0\% to 6\% by simulating the predictor $x$ from
$\mathcal{U}(-1,1)$ and the response $y$ is set the value -2 \citep{Nguyen2016}. An example of simulated samples with left-censoring
level 7.5\% form the MoE-SLap-CR, MoE-SBS-CR and MoE-SGIG-CR models and containing 6\% outliers is shown in Figure \ref{Sim4}.
In each trial of 500 replications, the MoE-N-CR, MoE-T-CR, MoE-CN-CR, and MoE-SL-CR models are fitted to the generated data.
Figure \ref{Sim42} shows an example fitted MoE curves to the data generated from the MoE-SLap-CR, MoE-SBS-CR and MoE-SGIG-CR models.
It can obviously be seen that the heavy-tailed models provide better fit and platforms for describing the data than the 
MoE-N-CR model. 

\begin{table*}[!tp]
	\centering{\caption{Simulation results for assessing the robustness of the proposed MoE model to outliers under various censoring levels and
			outliers percentages.\label{tab41}}
		\scalebox{0.7}{\hspace{-0.5cm}
			\begin{tabular}{cccccccccccccccccccccccccccccccccc}
				\hline
			 &Cens. Level $\rightarrow$& \multicolumn{4}{c}{7.5\%} && \multicolumn{4}{c}{30\%}\\
				\cline{3-6}\cline{8-11}
	True model & Fitted model & 0\% & 2\% & 4\% & 6\% && 0\% & 2\% & 4\% & 6\% \\
				\hline
				& MoE-N-CR & 0.0347 & 0.0948 & 0.1418 & 0.1897 && 0.1029 & 0.1626 & 0.2357 & 0.2776\\
	MoE-SGIG-CR & MoE-T-CR & 0.0297 & 0.0855 & 0.1195 & 0.1623 && 0.0698 & 0.1375 & 0.1857 & 0.2068 \\
				& MoE-CN-CR& 0.0334 & 0.0865 & 0.1232 & 0.1692 && 0.0928 & 0.1379 & 0.2013 & 0.2481\\
				& MoE-SL-CR& 0.0300 & 0.0856 & 0.1201 & 0.1642 && 0.0733 & 0.1392 & 0.1891 & 0.2161\\
				\\
				& MoE-N-CR & 0.0385 & 0.0979 & 0.1436 & 0.1936 && 0.1061 & 0.1657 & 0.2338 & 0.2788\\
	MoE-SBS-CR  & MoE-T-CR & 0.0342 & 0.0885 & 0.1226 & 0.1654 && 0.0735 & 0.1416 & 0.1918 & 0.2292\\
				& MoE-CN-CR& 0.0375 & 0.0897 & 0.1260 & 0.1729 && 0.0981 & 0.1427 & 0.2076 & 0.2507\\
				& MoE-SL-CR& 0.0344 & 0.0889 & 0.1230 & 0.1662 && 0.0789 & 0.1437 & 0.1932 & 0.2284\\
				\\
				& MoE-N-CR & 0.0451 & 0.1050 & 0.1512 & 0.1978 && 0.1142 & 0.1772 & 0.2401 & 0.2892\\
	MoE-SLap-CR  & MoE-T-CR & 0.0406 & 0.0950 & 0.1287 & 0.1712 && 0.0827 & 0.1519 & 0.1979 & 0.2289\\
				& MoE-CN-CR& 0.0437 & 0.0963 & 0.1329 & 0.1779 && 0.1032 & 0.1539 & 0.2165 & 0.2562\\
				& MoE-SL-CR& 0.0407 & 0.0953 & 0.1290 & 0.1719 && 0.0886 & 0.1548 & 0.2010 & 0.2394\\		   
				\hline
			\end{tabular}
	}}
\end{table*}
To assess the impact of the outliers on the parameter estimates and on the quality of the results, 
we calculate, in each 500 replications, the mean square error between the true regression mean function and the estimated one, defined as
\[
\text{MSE} = \frac{1}{500}\sum_{i=1}^{500}\Big(E_{\hat{\bm\Theta}}(\bm x_i,\bm r_i) - E_{\bm\Theta_{true}}(\bm x_i,\bm r_i)\Big)^2,
\]
where $E_{\bm\Theta}(\bm x_i,\bm r_i) = \sum_{j=1}^G \pi_j(\bm r_i;\bm\tau) \bm x_i^\top\bm\beta_j$ evaluated at 
the true and estimated parameters. Table \ref{tab41} shows, for each of the four MoE models, the average of MSE 
for an increasing percentage of outliers and censoring in the data.
First, one can see that the MSE tends towered zero as the level of censoring and outliers approach zeros for all cases of the MoE-SMN-CR model. Since the three considered scenarios generate fat-tailed data, it can be observed that 
without outliers (c = 0\%) the error of the MoE-N-CR model is greater than those of the other MoE models, reflecting its lack of 
robustness. Upon inspection of Table \ref{tab41}, one can conclude that by adding outliers to the data, the MoE-T-CR (and the MoE-SL-CR 
in the second order) model clearly outperforms others for all situations. It highlights that the  MoE-T-CR model is
much more robust to outliers under these data generating scenarios.

\section{Real data analysis}\label{sec5}
\begin{figure}[!t]
 	\begin{center}
 		\includegraphics[height=5cm, width=10cm]{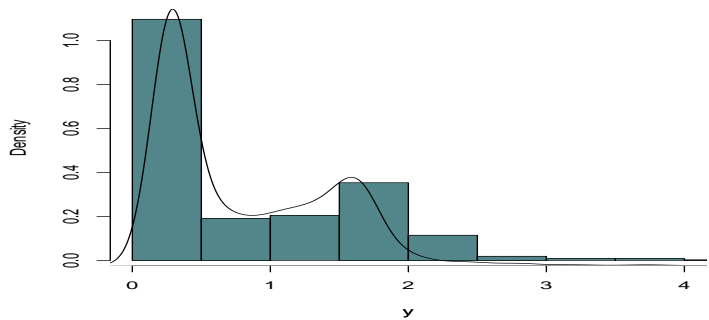}
 	\end{center}
 	\vspace{-0.7cm}
 	\caption{The histogram of the response variable $y$ overlaid with its Kernel density estimate.}\label{Wage1}
\end{figure}

This section considers the wage rates of 753 married women dataset, previously analyzed by \cite{mroz1987, 
	caudill2012, karlsson2014}, for illustrative purposes of the developed novel MoE-SMN-CR model. This dataset contains 753 observed wage rates 
(hours of working outside the home) of married white women between the ages of 30 and 60 in 1975, of whom 325 have zero hours working.
Recently, \cite{Zeller2018} reanalyzed the wage-rates dataset in order to illustrate the performance of the
FM of censored linear regression models based on the SMN class of distributions which is made available in 
the $\mathtt{R}$ package ``\textbf{CensMixReg}". Hereafter, we will denote the FM of censored linear regression models based on the
normal, Student-$t$, slash and contaminated-normal distributions
\citep{Zeller2018}, respectively by FM-N-CR, FM-T-CR, FM-SL-CR and FM-CN-CR. 
By considering the wife's annual work hours outside 
home scaled by 1000 as the response variable ($y$) which has 43.16\% level of left-censoring, and the explanatory 
variables including ($x_1$) the wife’s education in years, ($x_2$) the wife’s age, ($x_3$) the wife’s previous labor 
market experience and ($x_4$) the wife’s previous labor market experience squared, \cite{caudill2012, karlsson2014}
and \cite{Zeller2018} concluded that a mixture of two components linear regression censored model provides an
appropriate platform for analyzing this dataset. Figure \ref{Wage1} shows the histograms of $y$ overlaid with the estimated kernel 
density curve. The bimodality of the data and the suitability of the two-component mixture model to fit the data can 
be observed. It could be mentioned from the histogram that this is a heavily right-tailed data.  

The FM modeling allows clustering of the data in terms of the estimated (posterior) probability, $\hat{z}_{ij}$, that a 
single point belongs to a given group. Although the previous works on the wage-rates dataset focused on the aforementioned 
explanatory variables and showed that only these variables have significant effects on $y$, there are eleven measures that could provide
more information in investigating the complex relationship of random phenomena under study. One of those variable
that we will use for clustering purposes is the living status, labeled as ``city", that takes 1 for living in the city
and 0 for otherwise. Assuming ``city" as the group indicator, one can obtain the posterior probability $\hat{z}_{ij}$ 
and can therefore compute the clustering criteria MCR, RI, ARI and JCI of the FM regression models proposed by \cite{Zeller2018}.
In this regard, the posterior probabilities of the two-component FM-N-CR, FM-T-CR, FM-SL-CR and FM-CN-CR models are computed 
by fitting them to the considered data. It is observed that all of the models proposed by \cite{Zeller2018} assign data points
to one group.

As the advantages of the MoE model, it is possible for the investigator to choose some covariates 
for the gating function. In analyzing wage-rates data, we consider $\bm x=(1,x_1,x_2,x_3,x_4)^\top$ and $\bm r =(1,r_1,r_2,x_2)^\top$
for gating function, where ($r_1$) is the unemployment rate in county of residence and ($r_2$) is the number of 
kids less than 6 years old. We note that the covariates of the gating function can be the same as $\bm x$,
however by considering various combinations of the available explanatory variables, we observe that these three variables 
provide a better clustering performance. An interesting open issue for future work could be the variable selection problem
for both $\bm x$ and $\bm r$ in the MoE models.  

By fitting the MoE-N-CR, MoE-T-CR, MoE-SL-CR, and MoE-CN-CR models to these data for $G=1,\ldots,4$, 
the two-component MoE model has been selected based on the BIC. It should be noted that our results are not directly comparable 
with those obtained by \cite{karlsson2014} since they imposed some restrictions on $\bm\beta$ for estimation.
Moreover, it is clear that adding more variables to the model will definitely affect on the likelihood.
We therefore can not compare the results of model selection criteria, the AIC and BIC, with those reported by  
\cite{Zeller2018}. Table \ref{real2} shows the ML results obtained by fitting the four considered models.
The information-based approach for approximating standard error (SE) of parameter estimates are given in \ref{appB}.
We found that the estimated gating parameters are moderately significant, revealing that the considered 
covariates $\bm r$ have an effect on the analysis.
Results based on AIC and BIC indicate that the MoE-T-CR and MoE-SL-CR models provides an improved fit of the data over 
the other models. Moreover, by comparing the clustering criteria in Tables \ref{real2}, 
it turns out that the MoE-SL-CR models yields quite better classification. 

\begin{table*}[!t]
	\centering{\caption{ML estimates with corresponding approximate standard errors(SE) together with their AIC, BIC, 
			and clustering performance measures.\label{real2}}
		\scalebox{0.7}{\hspace{-0.5cm}
			\begin{tabular}{ccccccccccccccccc}
				\hline
				&& \multicolumn{2}{c}{MoE-N-CR} &&
				\multicolumn{2}{c}{MoE-T-CR} & & \multicolumn{2}{c}{MoE-SL-CR} & & \multicolumn{2}{c}{MoE-CN-CR}\\
				\cline{3-4}\cline{6-7}\cline{9-10}\cline{12-13}
				Parameter $\downarrow$ && Estimates & SE &&  Estimates & SE &&  Estimates & SE &&  Estimates & SE\\
				\hline 
				$\beta_{10}$ && 5.5476 & 0.6362 && 5.5438 & 0.6573 && 5.6223 & 0.7524 && 5.4714 & 0.9077\\
				$\beta_{11}$ &&-0.0554 & 0.0268 &&-0.0627 & 0.0027 &&-0.0658 & 0.0287 &&-0.0607 & 0.0227\\
				$\beta_{12}$ &&-0.1272 & 0.0130 &&-0.1256 & 0.0014 &&-0.1227 & 0.0167 &&-0.1212 & 0.0212\\
				$\beta_{13}$ && 0.0653 & 0.0355 && 0.0822 & 0.0050 && 0.0371 & 0.0063 && 0.0485 & 0.0114\\
				$\beta_{14}$ && 0.0004 & 0.0002 &&-0.0003 & 0.0001 && 0.0013 & 0.0029 && 0.0009 & 0.0007\\
				$\beta_{20}$ && 1.5064 & 0.2850 && 0.7306 & 0.0675 && 1.3579 & 0.4638 && 1.3405 & 0.2478\\
				$\beta_{21}$ && 0.0165 & 0.0025 && 0.0109 & 0.0025 && 0.0259 & 0.0051 && 0.0212 & 0.0028\\
				$\beta_{22}$ &&-0.0592 & 0.0125 &&-0.0410 & 0.0013 &&-0.0578 & 0.0109 &&-0.0560 & 0.0098\\
				$\beta_{23}$ && 0.2418 & 0.0205 && 0.2424 & 0.0018 && 0.2426 & 0.0207 && 0.2404 & 0.0128\\
				$\beta_{24}$ &&-0.0047 & 0.0006 &&-0.0049 & 0.0001 &&-0.0048 & 0.0007 &&-0.0047 & 0.0021\\
				$\sigma^2_{1}$ && 0.5001 & 0.0682 && 0.4365 & 0.0066 && 0.3773 & 0.0836 && 0.4173 & 0.1109\\
				$\sigma^2_{2}$ && 0.7130 & 0.0568 && 0.4661 & 0.0043 && 0.3120 & 0.0367 && 0.4214 & 0.1291\\
				$\nu_1$ && -- & -- && 9.3049 & -- && 9.0866 & -- && 0.0342&--\\
				$\nu_2$ && -- & -- && 6.2745 & -- && 1.8225 & -- && 0.1577&--\\
				$\gamma$ &&  -- & --  &&  -- & -- && -- & -- && 0.2643 &-- \\
				$\gamma$ &&  -- & --  &&  -- & -- && -- & -- && 0.2237 &-- \\
				$\tau_0$ &&26.7338 & 5.6234 &&48.3470 & 6.3394 &&14.4513 & 3.9352 && 17.4136 & 4.6459\\
				$\tau_1$ && 0.2519 & 0.1023 && 0.3414 & 0.2210 && 0.1845 & 0.0138 && 0.1999 & 0.0851\\
				$\tau_2$ && 4.1959 & 1.2368 && 6.9057 & 1.7441 && 0.8211 & 0.1362 && 0.7284 & 0.1246\\
				$\tau_3$ &&-0.7383 & 0.2304 &&-1.2943 & 0.3588 &&-0.4177 & 0.1394 &&-0.4912 & 0.1537\\
				\hline
				AIC && 1234.5830 &&& 1219.2230 &&& 1219.2830 &&& 1224.094 \\
				BIC && 1308.5680 &&& 1302.4570 &&& 1302.5160 &&& 1316.575 \\
				RI  && 0.5123   &&&  0.5214  &&&  0.5323  &&& 0.5118 \\ 
				JCI && 0.3676   &&&  0.3847  &&&  0.4029  &&& 0.3713 \\	 
				\hline
			\end{tabular}   
	}}
\end{table*}

\section{Conclusions and discussions}\label{sec6}
This paper proposed a new robust mixture of linear experts model for the censored data based 
on the scale-mixture of normal class of distributions. This MoE-SMN-CR model extended the classical 
MoE model which has been demonstrated to solve the two challenges to deal with heavy-tail 
distributed data and outliers as well as censored data. 
The newly proposed MoE-SMN-CR model is very extensive which extends the classical MoE model and includes
FM regression and FM regression for censored data proposed by \cite{Zeller2018} as special cases. 
The use of covariates in the gating function is an advantage of the MoE models which might result in 
better classification of the data. Utilizing the embedded hierarchical
structure of the SMN class of distributions, we developed an innovative EM-type algorithm to obtain ML parameter 
estimates computationally. We implemented this model in $\mathtt{R}$ and the computing program can be obtained from 
the authors upon request. 

Four Monte-Carlo simulation studies were conducted to investigate the performance of the model in applications 
both for non-linear regression and prediction and for model-based clustering. Results of simulation studies 
confirmed that the proposed MoE-SMN-CR model can provide evidence of the robustness to the outliers and atypical
observations. Finally, a real-world data analysis demonstrated the applicability and benefit of the proposed 
approach for practical applications. 

As discussed in Section \ref{sec5}, an interesting future direction of the current work is the variable selection for both
parts of the regression and gating function. The utility of our current approach can be further extended to the 
multiple regression on multivariate data rather than simple regression on univariate data, which we are actively exploring.
Another possible extension of the work herein is to consider a full Bayesian approach as a basis of inference and prediction 
\citep{peng1996,Zens2019}.  

\section*{Acknowledgments}
This work is based upon research supported by the South Africa National Research Foundation and South 
Africa Medical Research Council (South Africa DST-NRF-SAMRC SARChI Research Chair in Biostatistics, 
Grant number 114613), as well as by the National Research Foundation of South Africa (Grant Numbers 127727). Opinions expressed and conclusions arrived at are those of the author and are not necessarily to be attributed to the NRF.

\appendix
\def\theequation{A.\arabic{equation}}
\setcounter{equation}{0}

\section{Conditional expectations of the special cases of the SMN distributions}\label{appa}
{\bf Uncensored observations:} 
For the uncensored data $y_i$, we have $\rho_i =0$. Therefore, the only necessary conditional expectation
$\hat{u}_{ij}^{(k)}=E(U_{ij}|Y=y_i,\hat{\bm\theta}_j^{(k)})$ for the considered models can be computed as follows. 

\begin{itemize}
	\item[$\bullet$] If $Y\sim \mathcal{N}\big(\bm x_i^\top\hat{\bm{\beta}}_j^{(k)},\hat{\sigma}_j^{2(k)}\big)$, in this case, $U=1$ with probability one, and so $\hat{u}_{ij}^{(k)}=1$
	
	\item[$\bullet$] If $Y\sim \mathcal{T}\big(\bm x_i^\top\hat{\bm{\beta}}_j^{(k)},\hat{\sigma}_j^{2(k)},\hat{\nu}_j^{(k)}\big)$, We have
	\begin{equation*}
	\hat{u}_{ij}^{(k)}= \frac{\hat{\nu}_j^{(k)} + 1}{\hat{\nu}_j^{(k)} +\delta\Big(y_i, \bm x_i^\top\hat{\bm{\beta}}_j^{(k)},\hat{\sigma}_j^{(k)}\Big) },
	\end{equation*}
	where $\delta(y,\mu,\sigma) =\Big((y-\mu)/\sigma \Big)^2$.
	
	\item[$\bullet$] If $Y\sim \mathcal{SL}\big(\bm x_i^\top\hat{\bm{\beta}}_j^{(k)},\hat{\sigma}_j^{2(k)},\hat{\nu}_j^{(k)}\big)$, We have
	\begin{equation*}
	\hat{u}_{ij}^{(k)}= 2 \left(\delta\Big(y_i, \bm x_i^\top\hat{\bm{\beta}}_j^{(k)},\hat{\sigma}_j^{(k)}\Big)\right)^{-1} 
	\frac{\Gamma\left(\hat{\nu}_j^{(k)} + 1.5,\; 0.5 
		\delta\Big(y_i, x_i^\top\hat{\bm{\beta}}_j^{(k)},\hat{\sigma}_j^{(k)}\Big)\right) }{
		\Gamma\left(\hat{\nu}_j^{(k)}+ 0.5,\; 0.5  
		\delta\Big(y_i, \bm x_i^\top\hat{\bm{\beta}}_j^{(k)},\hat{\sigma}_j^{(k)}\Big)\right)}.
	\end{equation*}
	
	\item[$\bullet$] If $Y\sim \mathcal{CN}\big(\bm x_i^\top\hat{\bm{\beta}}_j^{(k)},\hat{\sigma}_j^{2(k)},\hat{\nu}_j^{(k)},\hat{\gamma}_j^{(k)}\big)$, We have
	\begin{equation*}
	\hat{u}_{ij}^{(k)}= \frac{1-\hat{\nu}_j^{(k)} + \hat{\nu}_j^{(k)} 
		(\hat{\gamma}_j^{(k)})^{1.5} \exp\Big\{0.5(1-\hat{\gamma}_j^{(k)}) 
		\delta\Big(y_i, \bm x_i^\top\hat{\bm{\beta}}_j^{(k)},\hat{\sigma}_j^{(k)}\Big)\Big\}}{
		1-\hat{\nu}_j^{(k)} + \hat{\nu}_j^{(k)} (\hat{\gamma}_j^{(k)})^{0.5} 
		\exp\Big\{0.5(1-\hat{\gamma}_j^{(k)}) \delta\Big(y_i, \bm x_i^\top\hat{\bm{\beta}}_j^{(k)},\hat{\sigma}_j^{(k)}\Big)\Big\}}.
	\end{equation*}
	
\end{itemize}

{\bf Censored cases:} 
In the censored cases, we have $\rho_i =1$. For the sake of notation, 
let 
$$T_{ij}^{(k)} = \dfrac{Y_i- \bm x_i^\top \hat{\bm{\beta}}_j^{(k)}}{\hat{\sigma}_j^{(k)}}\sim\text{SMN}(0,1,\hat{\bm\nu}_j^{(k)}), \qquad 
\hat{t}_{ij1}^{(k)}=\frac{c_{i1}- \bm x_i^\top \hat{\bm{\beta}}_j^{(k)}}{\hat{\sigma}_j^{(k)}}, \qquad
\hat{t}_{ij2}^{(k)}=\frac{c_{i2}- \bm x_i^\top \hat{\bm{\beta}}_j^{(k)}}{\hat{\sigma}_j^{(k)}}.  $$
Therefore, the necessary conditional expectations 
$\hat{u}_{ij}^{(k)}=E(U_{ij}|c_{i1} \leq Y_i \leq c_{i2},\hat{\bm\theta}_j^{(k)})$,
$\widehat{uy}_{ij}^{(k)}= E( U_{i}Y_i | c_{i1} \leq Y_i \leq c_{i2}, \hat{\bm\theta}_j^{(k)})$, and
$\widehat{uy^2}_{ij}^{(k)} = E( U_{i}Y_i^2 | c_{i1} \leq Y_i \leq c_{i2}, \hat{\bm\theta}_j^{(k)})$
for the considered models can be computed as follows.
\begin{equation*}
\hat{u}_{ij}^{(k)}= E\left( U_{ij} | \hat{t}_{ij1}^{(k)} \leq T_{ij}^{(k)} \leq \hat{t}_{ij2}^{(k)}, \hat{\bm\theta}_j^{(k)}\right) 
= \frac{E_\Phi\left(1,\hat{t}_{ij2}^{(k)}\right) - E_\Phi\left(1,\hat{t}_{ij1}^{(k)}\right) }{F_{SMN}\left( \hat{t}_{ij2}^{(k)};\hat{\bm\nu}_{j}^{(k)}\right) - F_{SMN}\left( \hat{t}_{ij1}^{(k)};\hat{\bm\nu}_{j}^{(k)}\right) },
\end{equation*}

\begin{align*}
\widehat{uy}_{ij}^{(k)} 
=& \left( x_i^\top \hat{\bm{\beta}}_j^{(k)}\right) \widehat{u}_{ij}^{(k)} + 
\hat{\sigma}_j^{(k)}  E\left( U_{ij} T_{ij}\Big| \hat{t}_{ij1}^{(k)} \leq T_{ij}^{(k)} \leq \hat{t}_{ij2}^{(k)}, \hat{\bm\theta}_j^{(k)} \right) \nonumber \\
= & \left( x_i^\top \hat{\bm{\beta}}_j^{(k)}\right) \left\lbrace  \frac{E_\Phi\left(1,\hat{t}_{ij2}^{(k)}\right) - E_\Phi\left(1,\hat{t}_{ij1}^{(k)}\right) }{F_{SMN}\left( \hat{t}_{ij2}^{(k)};\hat{\bm\nu}_{j}^{(k)}\right) - F_{SMN}\left( \hat{t}_{ij1}^{(k)};\hat{\bm\nu}_{j}^{(k)}\right) } \right\rbrace  +
\hat{\sigma}_j^{(k)} \left\lbrace  \frac{E_\phi \left(0.5,\hat{t}_{ij1}^{(k)}\right) - E_\phi\left(0.5,\hat{t}_{ij2}^{(k)}\right) }{F_{SMN}\left(\hat{t}_{ij2}^{(k)};\hat{\bm\nu}_{j}^{(k)} \right) - F_{SMN}\left(\hat{t}_{ij1}^{(k)};\hat{\bm\nu}_{j}^{(k)}\right)} \right\rbrace ,
\end{align*}

\begin{align*}
\widehat{uy^2}_{ij}^{(k)}
=& \left( x_i^\top \hat{\bm{\beta_j}}^{(k)}\right)^2 \widehat{u}_{ij}^{(k)} 
+ 2\left( x_i^\top \hat{\bm{\beta_j}}^{(k)}\right) \hat{\sigma}_j^{(k)} \widehat{uy}_{ij}^{(k)}
+\hat{\sigma}_j^{2(k)} E\left( U_{ij} T_{ij}^{2}\Big| \hat{t}_{ij1}^{(k)} \leq T_{ij}^{(k)} \leq \hat{t}_{ij2}^{(k)}, \hat{\bm\theta}_j^{(k)} \right), \nonumber \\
=& \left( x_i^\top \hat{\bm{\beta_j}}^{(k)}\right)^2 \widehat{u}_{ij}^{(k)} 
+ 2\left( x_i^\top \hat{\bm{\beta_j}}^{(k)}\right) \hat{\sigma}_j^{(k)} \widehat{uy}_{ij}^{(k)} + \frac{\hat{\sigma}_j^{2(k)}}{F_{SMN}\left( \hat{t}_{ij2}^{(k)};\hat{\bm\nu}_{j}^{(k)}\right) - F_{SMN}\left( \hat{t}_{ij1}^{(k)};\hat{\bm\nu}_{j}^{(k)}\right) } \nonumber \\
& \qquad \left( E_\Phi\left(0,\hat{t}_{ij2}^{(k)}\right) - 
E_\Phi\left(0,\hat{t}_{ij1}^{(k)}\right) + 
\left( \hat{t}_{ij1}^{(k)}\right) E_\phi\left(0.5,\hat{t}_{ij1}^{(k)}\right) 
-\left( \hat{t}_{ij2}^{(k)}\right) E_\phi\left(0.5,\hat{t}_{ij2}^{(k)}\right)  \right), 
\end{align*}

where
\[
E_\phi(r,h) = E\left(U^r \phi\big(h\sqrt{U}\big) \right)\quad \text{and} \quad 
E_\Phi(r,h) = E\left(U^r \Phi\big(h\sqrt{U}\big) \right).
\]
In the following, the closed forms of $E_\phi(r,h)$ and $E_\Phi(r,h)$ for the special cases of SMN class of distributions are presented.
\begin{itemize}
	\item[$\bullet$] For the normal distribution, we have
	\begin{align*}
	E_\phi(r,h) = \phi(h) \quad \text{and} \quad E_\Phi(r,h)=\Phi(h).
	\end{align*}
	
	\item[$\bullet$] In the case of Student-$t$ distribution, we have	
	\begin{align*}
	E_\phi(r,h) &= \frac{\Gamma\left(\dfrac{\hat{\nu}_j^{(k)}+2r}{2}\right) }{\sqrt{2\pi}\Gamma(\hat{\nu}_j^{(k)}/2)} 
	\left(\frac{\hat{\nu}_j^{(k)}}{2} \right)^{\dfrac{\hat{\nu}_j^{(k)}}{2}} \left(\frac{2}{h^2+\hat{\nu}_j^{(k)}} \right)^{\dfrac{\hat{\nu}_j^{(k)}+2r}{2}},\\
	E_\Phi(r,h) &= \Gamma\left( \frac{\hat{\nu}_j^{(k)}+2r}{2}\right) \left(\frac{2}{\hat{\nu}_j^{(k)}} \right)^r F_{PVII}\left(h;\hat{\nu}_j^{(k)}+2r, \hat{\nu}_j^{(k)} \right) \Big/ \Gamma(\frac{\hat{\nu}_j^{(k)}}{2}).
	\end{align*}
	where $F_{PVII}(\cdot;\nu,\delta)$ denotes the cdf of Pearson type $VII$ distribution.
	\item[$\bullet$] For the slash model, we have
	\begin{align*}
	E_\phi(r,h)= \frac{\hat{\nu}_j^{(k)}}{\sqrt{2\pi}} \left( \frac{2}{h^2} \right)^{\hat{\nu}_j^{(k)}+r}\Gamma(\hat{\nu}_j^{(k)}+r, \frac{h^2}{2} )
	\quad \text{and} \quad 
	E_\Phi(r,h) = \frac{\hat{\nu}_j^{(k)}}{\hat{\nu}_j^{(k)}+r} F_{SL}\left(h;\hat{\nu}_j^{(k)}+r \right).
	\end{align*}
	
	\item[$\bullet$] For the contaminated-normal distribution, we have
	\begin{align*}
	E_\phi(r,h) &=\left(\hat{\gamma}_j^{(k)}\right)^r \hat{\nu}_j^{(k)} \phi\left(h\sqrt{\hat{\gamma_j}^{(k)}}\right) 
	+\left(1-\hat{\nu}_j^{(k)}\right) \phi(h),\\
	E_\Phi(r,h) &=\left(\hat{\gamma_j}^{(k)}\right)^r F_{CN}\left(h;\hat{\nu}_j^{(k)},\hat{\gamma}_j^{(k)}\right)+ 
	\left(1-\left(\hat{\gamma}_j^{(k)}\right)^r\right) \left(1-\hat{\nu}_j^{(k)}\right) \Phi(h).
	\end{align*}	
\end{itemize}

\def\theequation{B.\arabic{equation}}
\setcounter{equation}{0}

\section{The hazard function plot of the normal distribution}\label{appab}

\begin{figure}[!ht]
	\begin{center}
		\includegraphics[height=5cm, width=5cm]{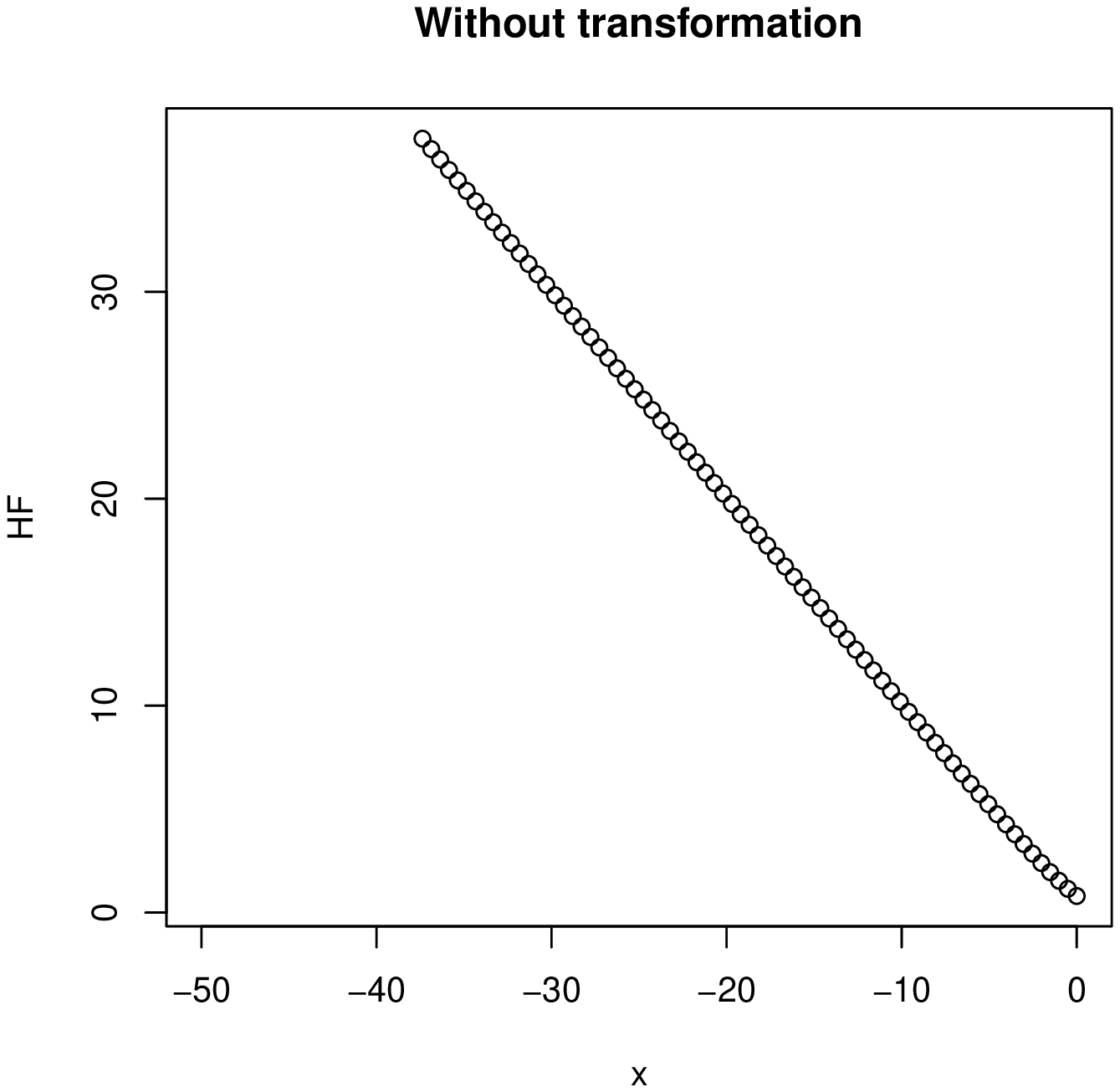}
		\includegraphics[height=5cm, width=5cm]{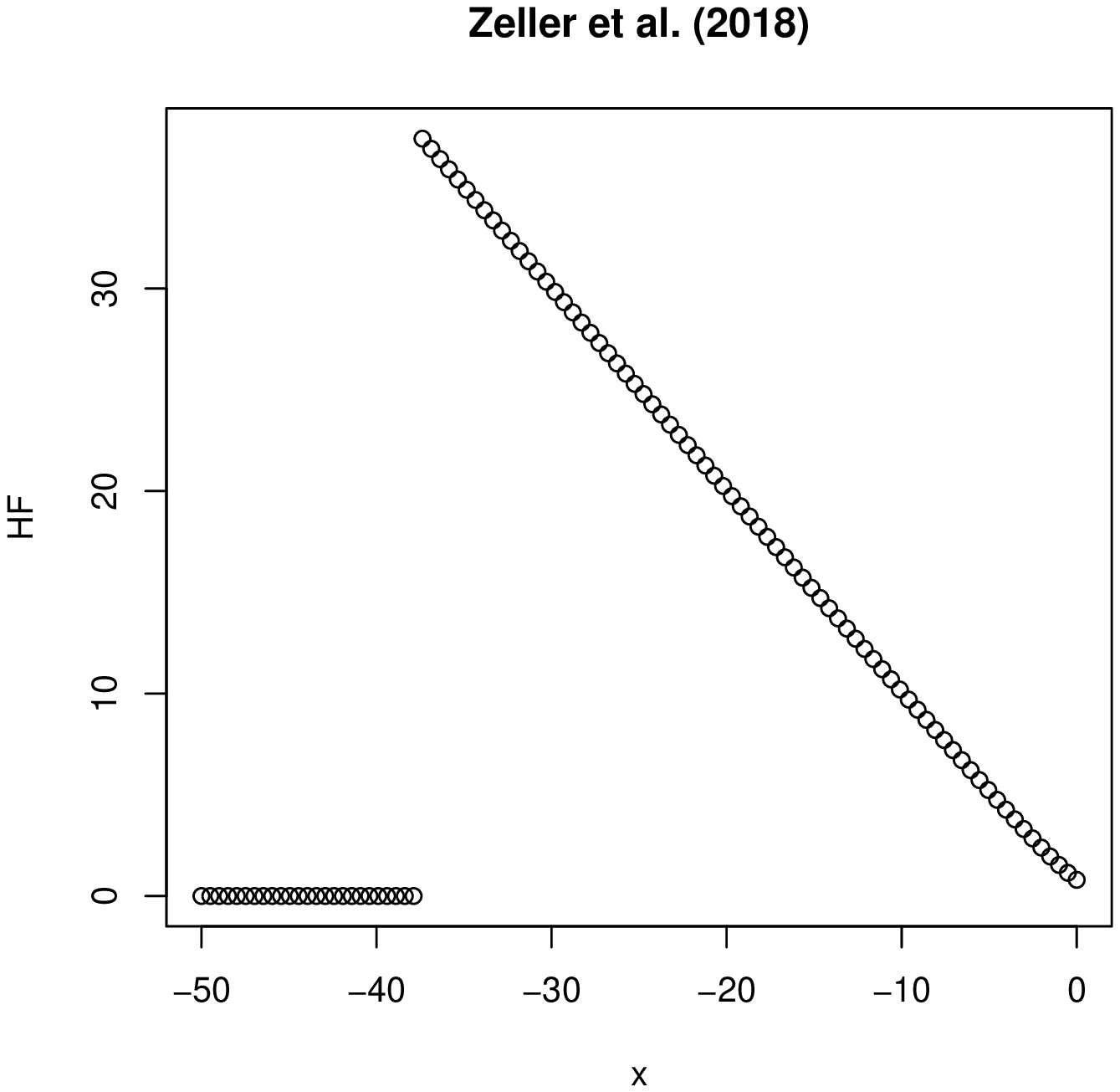}
		\includegraphics[height=5cm, width=5cm]{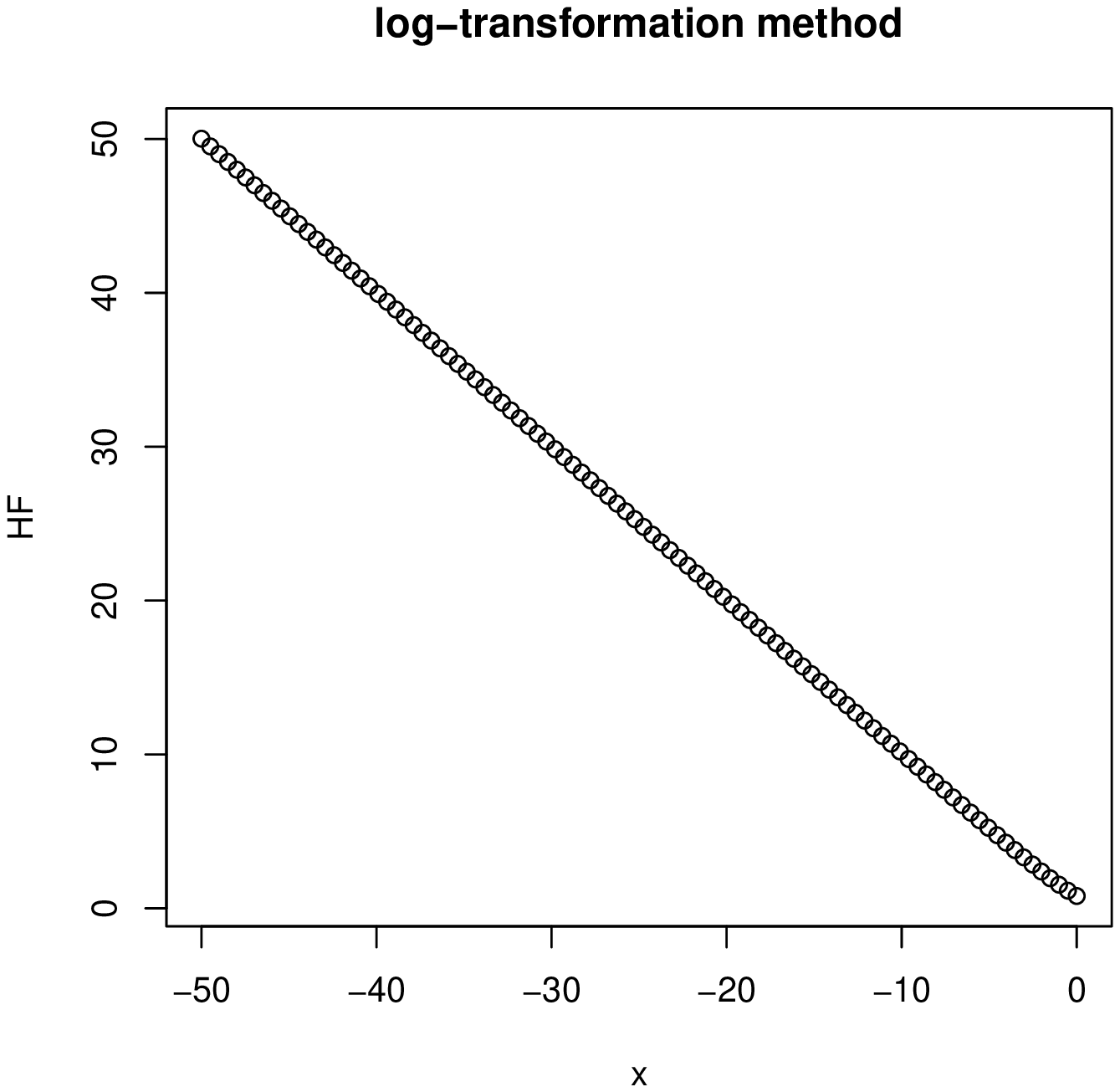}
	\end{center}
	\vspace{-0.5cm}
	\caption{The normal hazard function plots computed based on three ways in $\mathtt{R}$.}\label{HF}
\end{figure}

\def\theequation{C.\arabic{equation}}
\setcounter{equation}{0}

\section{Standard error estimates}\label{appB}
For estimating the standard error of the ML estimators, we follow \cite{meilijson1989} to exploit an information-based
method for calculating the asymptotic covariance matrix of the ML estimates. Let $\ell_{ci}$ 
be the
complete-data log-likelihood contributed from the $i$th observation. i.e.
\[
\ell_{ci}=\ell_c(\bm\Theta|\bm w_i^\top,\bm\rho_i^\top,\bm y_i^\top, \bm u_i^\top,\bm Z_i^\top)= \sum_{j=1}^G Z_{ij} \left\lbrace \log \pi_j(\bm r_i;\bm\tau)- \frac{1}{2}\log \sigma_j^2
- \frac{u_{i}}{2\sigma_j^2} (y_i-\bm x_i^\top\bm\beta_j)^2 +\log h(u_i;\bm \nu_j)\right\rbrace,
\]
Then, the Fisher information matrix can be approximated by
\begin{equation*}
	I_o(\hat{\Theta}|\bm y) = \sum_{i=1}^n\hat{\bm s}_i\hat{\bm s}_i^\top,
\end{equation*}
where $\hat{s}_i=E\Big(\dfrac{\partial\ell_{ci}}{\partial\bm\Theta}\; \Big|\; y_i,\hat{\bm\Theta}\Big)$
is the individual score vector corresponding to the $i$th observation. The elements of individual score vector $(\hat{s}_{i,\bm\tau_1}^\top,\ldots,\hat{s}_{i,\bm\tau_{G-1}}^\top\hat{s}_{i,\bm\beta_1}^\top,\ldots,
\hat{s}_{i,\bm\beta_{G}}^\top,\hat{s}_{i,\sigma_1^2},\ldots,\hat{s}_{i,\sigma_{G}^2})$ have the explicit forms as
\begin{align*}
	\hat{s}_{i,\bm\tau_j}&=E\Big(\dfrac{\partial\ell_{ci}}{\partial\bm\tau_j}\; \Big|\; y_i,\hat{\bm\Theta}\Big)
	=\left(\hat{z}_{ij}-\pi_j(\bm r_i;\hat{\bm\tau}) \right)\bm r_i, \qquad 
	\hat{s}_{i,\bm\beta_j}=E\Big(\dfrac{\partial\ell_{ci}}{\partial\bm\beta_j}\; \Big|\; y_i,\hat{\bm\Theta}\Big)
	=\frac{\hat{z}_{ij}}{\hat{\sigma}_j^2}\left(\widehat{uy}_{ij}\bm x_i-
	\hat{u}_{ij}\bm x_i^\top\hat{\bm\beta}_j\bm x_i \right) ,\\
	\hat{s}_{i,\sigma_j^2}&=E\Big(\dfrac{\partial\ell_{ci}}{\partial\sigma_j^2}\; \Big|\; y_i,\hat{\bm\Theta}\Big)
	=-\frac{\hat{z}_{ij}}{2\hat{\sigma}_j^4}\left(\hat{\sigma}_j^2-\widehat{uy^2}_{ij}-
	\hat{u}_{ij}(\bm x_i^\top\hat{\bm\beta}_j)^2 + 2 \widehat{uy}_{ij}\bm x_i^\top\hat{\bm\beta}_j  \right)  .
\end{align*}
As a result, the variance of the ML estimates can be consistently estimated from the diagonal of the
inverse of $I_o(\hat{\Theta}|\bm y)$ under some regularity conditions. We note that the standard error of $\hat{\nu}$ 
critically depends on the calculation of $E\big(\log(U_i)|y_i, \hat{\bm\Theta} \big)$ which is a computational challenge.
It could be mentioned that inverse of the $I_o(\hat{\Theta}|\bm y)$ is not always available. One can refer to \cite{yu2020}
to find an innovative interpolation procedure based on the cubic spline interpolation to directly estimate the asymptotic 
variance-covariance matrix of the ML estimates obtained by the EM algorithm.

\end{document}